\documentclass[10pt]{amsart}

\usepackage{stmaryrd}
\usepackage{amsfonts}
\usepackage{bbm}
\usepackage{mathrsfs}
\usepackage{latexsym,amssymb,amsmath,amscd,amsthm,amsxtra}
\usepackage[utf8]{inputenc}
\usepackage[T1]{fontenc}
\usepackage{nicefrac,mathtools,enumitem}
\usepackage{microtype}
\usepackage{changepage}

\newenvironment{pf}{\paragraph{{\sc Proof}}}{\qed\par\medskip}
\newtheorem{thm}{Theorem}[section]
\newtheorem{lem}[thm]{Lemma}

\newtheorem{pro}[thm]{Proposition}
\newtheorem{ex}[thm]{Example}
\newtheorem{rmk}[thm]{Remark}
\newtheorem{defi}[thm]{Definition}

\newcommand{\be }{\begin{equation}}
\newcommand{\ee }{\end{equation}}

\newcommand{\br}[1]{   [ \cdot,    \cdot  ]   }

\newcommand{\g}{\mathfrak g}

\newcommand{\id}{\mathbbm{i}}

\newcommand{\Ad}{\mathrm{Ad}}

\newcommand{\ad}{\mathrm{ad}}

\newcommand{\spr}{/\!\!/}

 \newcommand {\eps}{\epsilon}                  

\newcommand {\IR}{\mathbb{R}}

\newcommand {\R}{\mathbb R}

\newcommand {\sfPhi}{\mathsf{\Phi}}

\renewcommand {\ad}{\operatorname{ad}}
\renewcommand {\Ad}{\operatorname{Ad}}

\renewcommand {\id}{\operatorname{id}}

\newcommand{\U}{{\rm U}}

\newcommand{\Herm}{{\rm Herm}}

\newcommand {\wh}{\widehat}

\renewcommand {\t}{t}

\newcommand {\comment}[1]{\footnote{\textcolor{blue}{#1}}}
\renewcommand {\comment}[1]{}
\newcommand {\Omit}[1]{}

\newcommand{\Sym}{{\rm Sym}}

\begin{document}
\title[]{\small\bf
Stokes phenomenon, Gelfand-Zeitlin systems and relative Ginzburg-Weinstein linearization}

\author{\small XIAOMENG XU}

\date{}

\newcommand{\Addresses}{{
  \bigskip
  \footnotesize

  \textsc{DEPARTMENT OF MATHEMATICS, MASSACHUSETTS INSTITUTE OF TECHNOLOGY, CAMBRIDGE, MA 02139, USA}\par\nopagebreak
  \textit{Email address}: \texttt{xxu@mit.edu}

}}


\begin{abstract}
In 2007, Alekseev-Meinrenken proved that there exists a Ginzburg-Weinstein diffeomorphism from the dual Lie algebra ${\rm u}(n)^*$ to the dual Poisson Lie group $U(n)^*$ compatible with the Gelfand-Zeitlin integrable systems. In this paper, we explicitly construct such diffeomorphisms via Stokes phenomenon and Boalch's dual exponential maps. 
Then we introduce a relative version of the Ginzburg-Weinstein linearization motivated by irregular Riemann-Hilbert correspondence, and generalize the results of Enriquez-Etingof-Marshall to this relative setting. In particular, we prove the connection matrix for a certain irregular Riemann-Hilbert problem satisfies a relative gauge transformation equation of the Alekseev-Meinrenken dynamical r-matrices. This gauge equation is then derived as the semiclassical limit of the relative Drinfeld twist equation. 
\end{abstract}

\maketitle

\setcounter{tocdepth}{1}
\tableofcontents

\section{Introduction and main results}
The Ginzburg-Weinstein linearization theorem \cite{GW} states that for any compact Lie group $K$ with its standard Poisson structure, the dual Poisson Lie group $K^*$ is Poisson isomorphic to the dual of the Lie algebra $\mathfrak{k}^*$, with its canonical linear (Kostant-Kirillov-Souriau) Poisson structure. When $K=U(n)$, the Poisson manifolds ${\rm u}(n)^*$ and $U(n)^*$ carry more structures: Guillemin-Sternberg \cite{GS1} introduced the Gelfand–Zeitlin integrable system on ${\rm u}(n)^*$; later on, Flaschka-Ratiu \cite{FR} described a multiplicative Gelfand-Zeitlin system for the dual Poisson Lie group $U(n)^*$. Then it was proved by Alekseev-Meinrenken \cite{AM} that there exists a Ginzburg-Weinstein linearization which intertwines the Gelfand-Zeitlin systems on ${\rm u}(n)^*$ and $U(n)^*$.

There are various generalizations of Ginzburg-Weinstein linearization to the complex and formal setting.
In \cite{Boalch1}, Boalch pointed out that, for $G={\rm GL}_n(\mathbb{C})$ equipped with the standard Poisson Lie group structure, the dual Poisson Lie group $G^*$ is identified with a moduli space of meromorphic connections with certain irregular singularity. This viewpoint enabled him
to define a class of "dual exponential maps" $\nu:\g^*\to G^*$ by taking the Stokes data of the meromorphic connections. Then his remarkable result shows that these (irregular Riemann-Hilbert) maps $\nu:\g^*\rightarrow G^*$ are local Poisson isomorphisms. Later on in \cite{BoalchG}, this result, along with the definition of such Stokes data, was extended beyond ${\rm GL}_n(\mathbb{C})$ to any complex reductive Lie groups. On the other hand, Enriquez-Etingof-Marshall \cite{EEM} constructed formal Poisson isomorphisms between the formal Poisson manifolds $\mathfrak g^*$ and $G^*$. Their result relies on constructing a formal map $\rho:\mathfrak g^*\rightarrow G$ satisfying a vertex-IRF gauge transformation equation \cite{EN} of the Alekseev-Meinrenken r-matrix \cite{AM0}. Furthermore, one solution of this gauge equation is derived as the semiclassical limit of a Drinfeld twist. Ginzburg-Weinstein linearization in more general setting can be found in \cite{AM1}.

The comparison of the above two approaches enables us to unveil some unexpected relations between Stokes phenomenon, dynamical r-matrices and Drinfeld twists \cite{Xu} . These mysterious relations are later on understood in our joint work with Toledano Laredo \cite{TLXu} by studying the Stokes phenomenon of dynamical Knizhnik-Zamolodchikov equations \cite{FMTV}.
However, the possible role of Gelfand-Zeitlin systems in either Boalch's or Enriquez-Etingof-Marshall's approach was still unknown. 
\\

In this paper, we point out and then draw several consequences of a relation between Stokes phenomenon, Gelfand-Zeitlin (GZ) systems and relative Drinfeld twists. In particular, we will relate the
GZ systems to Boalch's dual exponential maps. In some sense, this gives a "moduli theoretic" interpretation of the multiplicative GZ
system, i.e., in terms of moduli spaces of meromorphic connections. In the following, we state our main results.

Let $G$ be a complex reductive Lie group with ${\rm Lie}(G)=\g$, and $\mathfrak t\subset \g$ a Cartan subalgebra. Let us consider the meromorphic connection on the trivial holomorphic principal $G$-bundle $P$ on $\mathbb{P}^1$ which has the form
\begin{eqnarray*}
\nabla=d-(\frac{\lambda}{z^2}+\frac{1}{2\pi i}\frac{A}{z})dz
\end{eqnarray*}
where $\lambda,A\in \mathfrak g$. We assume that $\lambda\in \mathfrak t$, and once fixed, the only variable is $A\in\g$. Given an initial Stokes sector (and a branch of ${\rm log}(z)$ on it), we consider the monodromy of $\nabla$ from $0$ to $\infty$, known as the connection matrix $C(A)\in G$ of $\nabla$, which is the ratio of two canonical solutions of $\nabla F=0$, one is around $\infty$ and another is on the Stokes sector at $0$. Varying $A\in\g$, we thus get a (densely defined) map $C:\g\rightarrow G$ by mapping $A\in\mathfrak g$ to the connection matrix $C(A)$ of $\nabla$. We call it the connection map associated to the irregular type $\frac{\lambda}{z}$ (and the initial Stokes sector). See Section \ref{defineC} for more details.

\subsection*{Connection maps and Gelfand-Zeitlin systems}
For each $0<k\le n$, let $\g={\rm gl}_k(\mathbb{C})$, and $\lambda_k\in\frak t$ (the set of diagonal matrices) whose centralizer is the Levi subalgebra $\mathfrak h={\rm gl}_{k-1}(\mathbb{C})+\mathfrak t\subset {\rm gl}_k(\mathbb{C})$. Here ${\rm gl}_{k-1}(\mathbb{C})\subset {\rm gl}_{k}(\mathbb{C})$ denotes the set of $(k-1)$th principal submatrices. Let $C_k:{\rm gl}_k(\mathbb{C})\rightarrow {\rm GL}_k(\mathbb{C})$ be the connection map associated to the irregular type $\frac{\lambda_k}{z}$. We denote by $C:=C_1\cdot\cdot\cdot C_n$ the pointwise multiplication of the connection matrix maps $C_k$. Then in Section \ref{defineC}, we prove
\begin{thm}
The map $\Gamma:={\rm Ad_C}\circ{\rm exp}:\Herm(n)\cong {\rm u}(n)^*\rightarrow \Herm^+(n)\cong U(n)^*$ is a Poisson diffeomorphism compatible with the Gelfand-Zeitlin systems.
\end{thm}
Here $\Herm(n)$ ($\Herm^+(n)$) denotes the set of (positive definite) Hermitian $n$ by $n$ matrices, which is naturally isomorphic to the dual Lie algebra ${
\rm u}(n)^*$ (dual Poisson Lie group $U(n)^*$). See Section \ref{RelatoPoisson} for more details. In this procedure, we actually break the Ginzburg-Weinstein linearization into $n$ "universal" steps, and each step is a relative linearization by a connection map $C_k$. This will become clear in Section \ref{Symdynamical}.

\subsection*{Irregular Riemann-Hilbert maps and relative Ginzburg-Weinstein linearization}
To relate the above result to symplectic geometry, we consider an extended moduli space, which is the set of isomorphism classes of triples $(P,\nabla,{\bf g})$, consisting of a connection $\nabla$ on a holomorphic trivial $G$-principal bundle $P$ with irregular type $\frac{\lambda}{z}$ and compatible framing ${\bf g}$, see Section \ref{IrrRHcorrs}. We assume the centralizer of $\lambda\in\mathfrak t$ is a Levi subalgebra $\mathfrak h=\mathfrak g_1+\mathfrak t$ with the semisimple subalgebra $\g_1\subset \mathfrak h\subset\g$.

The space of Stokes/monodromy data of the triples $(P,\nabla,{\bf g})$ inherits a symplectic structure via an irregular analogue of the Atiyah-Bott construction \cite{Boalch2, Boalch3}. It is isomorphic to a symplectic "slice" $(G\times \mathfrak t',\pi_\g)$ of the Lu-Weinstein symplectic double \cite{LW2}. Here $\mathfrak t'$ is the complement of the affine root hyperplanes: $\mathfrak t':= \{\Lambda\in \mathfrak t~|~ \alpha(\Lambda)\notin 2\pi i \mathbb{Z}\}$. 
On the other hand, the moduli space of the triples $(P,\nabla,{\bf g})$ is isomorphic to $G\times \mathfrak t'$, on which we introduce a natural symplectic structure $\pi_{\g_1}$ depending on the pair $\g_1\subset \g$.

Let $C_{\g_1}:\g^*\rightarrow G$ be the connection map associated to the irregular type $\frac{\lambda}{z}$. Then in Section \ref{IrrRHcorrs}, we prove that
\begin{thm}
The irregular Riemann-Hilbert map
$$\nu_{C_{\g_1}}:(G\times \mathfrak t',\pi_{\g_1})\rightarrow (G\times \mathfrak t',\pi_\g); \ (g,t)=(C_{\g_1}({\rm Ad}^*_gt)g, t),$$
associating the monodromy data to any triple $(P,\nabla,{\bf g})$ is a local symplectic isomorphism.
\end{thm}
This is analogue to the result in \cite{Boalch2,Boalch3,Xu} except here we drop the assumption of $\lambda$ being regular. This key feature of irregular Riemann-Hilbert correspondence motivates us to introduce a relative version of Ginzburg-Weinstein linearization (with respect to $\g_1\subset \g$), which is an $H$-equivariant (local) symplectic isomorphism from $(G\times \mathfrak t',\pi_{\g_1})$ to $(G\times \mathfrak t',\pi_\g)$. Here the Lie subgroup $H\subset G$, the integration of $\mathfrak h\subset \g$, acts on the first component of $G\times \mathfrak t'$ by left translation. In particular, the irregular Riemann-Hilbert map gives rise to such a relative linearization.

In the following, we give another approach to relative Ginzburg-Weinstein linearization via dynamical r-matrices and the theory of quantization of Lie bialgebras, which generalizes various results in \cite{EEM} to a relative setting.

\subsection*{Relation to dynamical r-matrices}
We introduce a relative gauge transformation equation for a map $\rho \in {\rm Map}(\g^*,G)$,
\begin{eqnarray*}
{r_{\scriptscriptstyle \rm AM}}_\g-(\otimes^2 {\rm Ad}_{\rho^{-1}})(r_{\scriptscriptstyle AM_{\g_1}})=(r_\g-r_{\g_1})^\rho,
\end{eqnarray*}
where $r_\g$ (resp. $r_{\g_1}$) and ${r_{\scriptscriptstyle \rm AM}}_\g$ (resp. ${r_{\scriptscriptstyle \rm AM}}_{\g_1}$) are the standard classical $r$-matrix and the Alekseev-Meinrenken dynamical r-matrix \cite{AM0} of $\g$ (resp. $\g_1$). See Section \ref{sectiongauge} for the conventions. Its relation with symplectic geometry is as follows. 

\begin{thm}
A map $\rho$ is an $H$-equivariant solution of the gauge equation if and only if
$$\nu_{\rho}:(G\times \mathfrak t',\pi_{\g_1})\rightarrow (G\times \mathfrak t',\pi_\g); \ (g,t)=(\rho ({\rm Ad}^*_gt)g, t),$$
is a relative Ginzburg-Weinstein linearization with respect to $\g_1\subset \g$.
\end{thm}
As an immediate consequence, the connection map $C_{\g_1}$ gives rise to an $H$-equivariant solution of the above gauge equation. 

\subsection*{Relation to quantization of Lie bialgebras}
Another construction of relative linearization is given via the theory of quantization of Lie bialgebras. Let $\Phi_g$ (resp. $\Phi_{\g_1}$) be an admissible associator of $\g$ (resp. $\g_1$) \cite{EH}. Let $J\in (U(\g)^{\otimes{2}}\llbracket\hbar\rrbracket)^\mathfrak h$ be a relative Drinfeld twist (see e.g. \cite{TL0}), i.e., $J$ satisfies the identity
\[\Phi_\g = (J^{2,3}J^{1,23})^{-1}\Phi_{\g_1} J^{1,2} J^{12,3}.\] 
Let us assume $J$ is admissible, and denote its semiclassical limit by $\rho:\g^*\rightarrow G$ (see Section \ref{relativetwist}).
\begin{pro}
The map $\rho$ is an $H$-equivariant formal solution of the equation \eqref{gaugeequation}. 
\end{pro}
Therefore, the semiclassical limit of a relative Drinfeld twist gives rise to a formal relative Ginzburg-Weinstein linearization. 

One natural question inspired by the above proposition is if there exists a relative Drinfeld twist whose semiclassical limit is $C_{\g_1}$. Such a relative twist may be constructed as a (quantum) connection matrix of the dynmaical Knizhnik–Zamolodchikov equation \cite{FMTV} by generalizing the construction in \cite{TL} to non-regular $\lambda$ case, and then the (quantum) connection matrix can be shown to be a quantization of the map $C_{\g_1}$ following the idea in \cite{TLXu}. 

\subsection*{Other related problems}
It is interesting to consider the isomonodromy deformation problem of the meromorphic connection $\nabla$ with irregular type $\frac{\lambda}{z}$, for those $\lambda$ with a fixed centralizer. This is supposed to recovery the quantum Weyl group action on the Poisson Lie group $G^*$ in the spirit of Boalch \cite{BoalchG}. 

In \cite{KW}, Kostant and Wallach introduced a complex version of the Gelfand-Zeitlin integrable system. It is interesting to generalize our result to that case.

\subsection*{Acknowledgements}
\noindent
I would like to thank Anton Alekseev, Philip Boalch, Pavel Etingof, Jianghua Lu and Valerio Toledano Laredo for their useful discussions and suggestions on this paper. This work is supported by the SNSF grant P2GEP2-165118.

\section{Gelfand-Zeitlin systems}\label{sec:intro}
\subsection{Gelfand-Zeitlin maps} Let $\Herm(n)$ denote the space of complex Hermitian $n\times n$-matrices. For $k\le n$ let $A^{(k)}\in \Herm(k)$ denote the $k$th 
principal submatrix (upper left $k\times k$ corner) of $A\in \Herm(n)$,
and $\tau^{(k)}_i(A)$-its ordered set of
eigenvalues, $\tau_1^{(k)}(A)\le \cdots\le \tau_k^{(k)}(A)$. 
The map 
\begin{equation}\label{eq:momentmap}
\tau\colon \Herm(n)\to \R^{\frac{n(n+1)}{2}},
\end{equation}
taking $A$ to the collection of numbers $\tau_i^{(k)}(A)$ for $1\le i\le
k\le n$, is a continuous map called the Gelfand-Zeitlin map.
Its image $\mathcal{C}(n)$ is the Gelfand-Zeitlin cone, cut out by the following inequalities,
\begin{equation}\label{eq:cone}
\tau_i^{(k+1)}\le \tau_i^{(k)}\le \tau_{i+1}^{(k+1)},\ \ 1\le
i\le k\le n-1.
\end{equation}
Let $\Herm^+(n)\subset \Herm(n)$ denote the subset of positive definite
Hermitian matrices, and define a logarithmic Gelfand-Zeitlin map 
\begin{equation}\label{eq:logmomentmap}
\mu\colon \Herm^+(n)\to \R^{\frac{n(n+1)}{2}},
\end{equation}
taking $A$ to the collection of numbers
$\mu^{(k)}_i(A)=\log(\tau^{(k)}_i(A))$. Then $\mu$ is a continuous
map from $\Herm^+(n)$ onto $\mathcal{C}(n)$.

\subsection{Gelfand-Zeitlin torus actions}\label{GZtorus} 
Let $\mathcal{C}_0(n)\subset \mathcal{C}(n)$ denote the subset where all of the eigenvalue inequalities
\eqref{eq:cone} are strict. Let $\Herm_0(n):=\tau^{-1}(\mathcal{C}_0(n))$ be the corresponding dense open subset of $\Herm(n)$. The $k$-torus 
$T(k)\subset U(k)$ of diagonal matrices acts on 
$\Herm_0(n)$ as follows,  
\begin{equation}\label{eq:taction}
 t\bullet A=\Ad_{U^{-1} t U}A,\ \ \ \ t\in T(k),\ A\in \Herm_0(n).
\end{equation}
Here $U\in U(k)\subset U(n)$ is a unitary matrix such that $\Ad_{U}A^{(k)}$ is
diagonal, with entries $\tau_1^{(k)},\ldots,\tau_k^{(k)}$. The action
is well-defined since $U^{-1}t U$ does not depend on the choice of
$U$, and preserves the Gelfand-Zeitlin map \eqref{eq:momentmap}.  The actions
of the various $T(k)$'s commute, hence they define an action of the Gelfand-Zeitlin torus
\[ T(n-1)\times \cdots \times T(1)\cong \U(1)^{(n-1)n/2}.\]
Here the torus $T(n)$ is excluded, since the 
action \eqref{eq:taction} is trivial for $k=n$.  

\subsection{Diffeomorphism compatible with Gelfand-Zeitlin systems}\label{linearalgebra}

In \cite{AM}, Alekseev and Meinrenken proved that there exists a diffeomorphism from $\Herm(n)$ to $\Herm^+(n)$, which intertwines the Gelfand-Zeitlin maps and actions (on $\Herm_0(n)$ and $\Herm^+_0(n)$). We will construct such diffeomorphisms via Stokes phenomenon. It relies on the following linear algebra result.

For each $0<k\le n$, let $C_k:\Herm(k)\rightarrow {\rm SU}(k)$ be a smooth map satisfying the conditions
\begin{enumerate}
\item[(a)]\label{it:a}  $C_k$ is a ${\rm SU}(k-1)$-equivariant map, i.e., $C_k(gAg^{-1})={\rm Ad}_gC_k(A)$, for any $g\in {\rm SU}(k-1)$;
\item[(b)]\label{it:b} for any $A\in \Herm(n)$, there is a block $\rm LU$ decomposition of $C_k(A) e^A C_k(A)^{-1}$ taking the form $$C_k(A) e^A C_k(A)^{-1}=\left(\begin{array}{cc}
{\rm Id}&0\\
\bar{b}_-& 1
\end{array} \right)\left(\begin{array}{cc}
e^{A^{(k-1)}} & 0\\
0& \star
\end{array} \right)\left(\begin{array}{cc}
{\rm Id} & \bar{b}_+\\
0& 1
\end{array} \right).$$
\end{enumerate}
We will think of $C_k:\Herm(k)\rightarrow {\rm SU}(k)$ as a map from $\Herm(n)$ to ${\rm SU}(n)$ using the projection of $\Herm(n)$ onto $\Herm(k)$ and the natural group homomorphism $\iota:{\rm SU}(k)\rightarrow {\rm SU}(n)$. That is for any $A\in \Herm(n)$, $C_k(A):=\iota (C_k(A^{(k)}))$. Then we have

\begin{thm}\label{th:GWiso}
Let $C:=C_1\cdot\cdot\cdot C_n$ be the map from $\Herm(n)$ to ${\rm SU}(n)$ given by the pointwise multiplication. Then $\Gamma:={\rm Ad}_C\circ {\rm exp}$ is a diffeomorphism 
\[ \Gamma\colon \Herm(n)\to
  \Herm^+(n)\]
such that 
\begin{itemize}
\item $\Gamma$ intertwines the
  Gelfand-Zeitlin maps: $\mu\circ \Gamma=\tau$.
\item $\Gamma$ intertwines the
  Gelfand-Zeitlin torus actions on $\Herm_0(n)$ and $\Herm_0^+(n)$. 
\end{itemize}
\end{thm}
\pf We will prove this theorem inductively on $n$. When $n=1$, $C=1$, the result is obvious. 

For the inductive step $n>1$, we assume $\phi:=C_1\cdot\cdot\cdot C_{n-1}$ is such that the map ${\rm Ad}_\phi\circ {\rm exp}:\Herm(n-1)\to
  \Herm^+(n-1)$ intertwines the
  Gelfand-Zeitlin maps and torus actions. 
  
Let us consider the map ${\rm Ad}_{\phi C_n}\circ {\rm exp}:\Herm(n)\to
  \Herm^+(n)$. By condition $(a)$, i.e., the ${\rm SU}(n-1)$-equivariance of $C_n$, we have for any $A\in \Herm(n)$
  $${\rm Ad}_{\phi C_n}(e^A)=C_n(\phi A\phi^{-1})e^{\phi A\phi^{-1}}C_n(\phi A\phi^{-1})^{-1}.$$
On the other hand, according to the condition $(b)$ of $C_n$, we have 
$${\rm Ad}_{C_n(\phi A\phi^{-1})}(\phi e^A \phi^{-1})=\left(\begin{array}{cc}
{\rm Id}&0\\
\bar{b}_-& 1
\end{array} \right)\left(\begin{array}{cc}
\phi e^{A^{(k-1)}}\phi^{-1} & 0\\
0& \star
\end{array} \right)\left(\begin{array}{cc}
{\rm Id} & \bar{b}_+\\
0& 1
\end{array} \right).$$  
Here we use the fact $\phi$ is valued in ${\rm SU}(n-1)\subset {\rm SU}(n)$, and thus the $(n-1)$th principal submatrix $(\phi A\phi^{-1})^{(n-1)}=\phi A^{(n-1)}\phi^{-1}$. We have shown that the map ${\rm Ad}_{\phi C_n}\circ {\rm exp}$ takes the form
$$ \ \ A\mapsto \left(\begin{array}{cc}
\phi e^{A^{(k-1)}}\phi^{-1} & \star\\
\star & \star
\end{array} \right),$$
where the right hand side matrix (conjugate to $e^A$) has same eigenvalues as $e^A$. Hence by the inductive assumption about the map ${\rm Ad}_\phi\circ {\rm exp}$, we obtain that ${\rm Ad}_{\phi C_n}\circ{\rm exp}$ intertwines the
Gelfand-Zeitlin maps. 

For the Gelfand-Zeitlin torus action, recall that 
$T(k)\subset U(k)$ act on 
$\Herm_0(n)$ by $t\bullet A=\Ad_{U^{-1} t U}A$, where $k<n$ and $U\in U(k)\subset U(n)$ is a unitary matrix such that $\Ad_{U}A^{(k)}$ is
diagonal as section \ref{GZtorus}. We first prove that $\phi(t\bullet A)e^{t\bullet A}\phi(t\bullet A)^{-1}=t\bullet \left( \phi(A) e^A\phi(A)^{-1}\right)$ for any $t\in T(k)$ and $A\in \Herm_0(n)$, i.e., the map ${\rm Ad}_\phi\circ {\rm exp}:\Herm (n)\rightarrow \Herm^+(n)$ is $T(k)$-equivariant. 

Set $B=A^{(n-1)}\in \Herm_0(n-1)\subset \Herm_0(n)$. By the inductive assumption, we have $$\phi(t\bullet B)e^{t\bullet B}\phi(t\bullet B)^{-1}=t\bullet \left( \phi(B) e^{B}\phi(B)^{-1}\right).$$
This is to say the adjoint actions of $\phi(t\bullet B)U^{-1} t U$ and ${U'}^{-1} t U'\phi(B)$ on $e^{B}$ coincide, where $U'\in U(k)\subset U(n)$ is a matrix diagonalizing the $k$th principal submatrix of $\phi(B) e^{B}\phi(B)^{-1}$. We can actually choose $U'$ such that
\begin{eqnarray}\label{tcomm}\phi(t\bullet B)U^{-1} t U={U'}^{-1} t U'\phi(B).
\end{eqnarray}
This can be seen as follows: when $k=n-1$, by assumption $U\in U(n-1)$ diagonalizes the matrix $B$. Therefore, we have $t\bullet B=B$, and $U'$ can be chosen as $U\phi(B)^{-1}$. Identity \eqref{tcomm} follows; if $k<n-1$, let us write $\phi=\phi_k\cdot C_{k+1}\cdot\cdot\cdot C_{n-1}$, where $\phi_k:=C_1\cdot\cdot\cdot C_k$ is the pointwise multiplication of the first $k$ maps. Because $C_{i}$ is $U(k)$-equivariant for $i>k$, we have $$C_{k+1}(t\bullet B)\cdot\cdot\cdot C_{n-1}(t\bullet B)=U^{-1}tU C_{k+1}(B)\cdot\cdot\cdot C_{n-1}(B) U^{-1}t^{-1}U. $$Then identity \eqref{tcomm} is equivalent to $\phi_k(t\bullet B)U^{-1} t U={U'}^{-1} t U'\phi_k(B)$, which reduces to the first case. That is we have $\phi_k(t\bullet B)=\phi_k(B)$, and thus $U'$ can be chosen as $U\phi_k(B)^{-1}\subset U(k)$. Here we use the convention $\phi_k(B)=\phi_k(B^{(k)})$.

Recall that $B=A^{(n-1)}$. By definition $\phi(t\bullet A)$ only depends on $(t\bullet A)^{(n-1)}=t\bullet B$. That is $\phi(t\bullet A)=\phi(t\bullet B)$.
Therefore we have $$\phi(t\bullet A)e^{t\bullet A}\phi(t\bullet A)^{-1}=\Ad_{\phi(t\bullet B)U^{-1} t U}A=t\bullet \left( \phi(A) e^A\phi(A)^{-1}\right).$$ Here we use \eqref{tcomm} and the fact $\phi(A)=\phi(B)$ in the second identity. That is the map ${\rm Ad}_\phi\circ {\rm exp}$ is $T(k)$-equivariant.
Now because of the $T(k)$-equivarience of $C_n$ for $k<n$, the $T(k)$-equivarience of the map $\Gamma={\rm Ad}_{\phi\cdot C_n}\circ {\rm exp}={\rm Ad}_{C_n}\circ ({\rm Ad}_\phi\circ {\rm exp})$ becomes apparent. That is $\Gamma$ intertwines the
Gelfand-Zeitlin torus actions. 
\qed\\

A set of maps $C_k$ satisfying the condition $(a)$ and $(b)$ will be constructed via certain irregular Riemann-Hilbert problem in the next section.

\begin{rmk} According to Duistermaat \cite{Duistermaat}, for a real semi-simple Lie group $G$ with
Cartan decomposition $G=KP$, there exits a smooth map $\psi\colon \mathfrak {p}\to K$ such that
$\Gamma=\exp\circ \Ad_\psi\colon\mathfrak {p}\to P$ intertwines the 
`diagonal projection' with the `Iwasawa projection'. In \cite{Boalch1}, Boalch showed that connection maps for certain irregular Riemann-Hilbert problem give examples of Duistermaat maps.
\end{rmk}

{\bf Relation to the Alekseev-Meinrenken diffeomorphism.}
Let $\Sym(n)$ denote the space of symmetric $n\times n$-matrices, and $\Sym^+(n)$ its subset of positive definite matrices. In a similar way, we define surjective maps
\[\tau: \Sym(n)\rightarrow \mathcal{C}(n), \ \ \mu:\Sym^+(n)\rightarrow \mathcal{C}(n)\] in terms of eigenvalues of principal submatrices.
According to \cite{AM}, the restriction of the Gelfand-Zeitlin map to $\Herm_0(n)$ defines a
  principal bundle $\tau\colon \Herm_0(n)\to \mathcal{C}_0(n)$
  with structure group the Gelfand-Zeitlin torus. It further restricts to a principal bundle 
$\tau\colon \Sym_0(n) \to \mathcal{C}_0(n)$
with a discrete structure group $T_\R(n-1)\times\cdots \times T_\R(1)\cong (\mathbb{Z}_2)^{n(n-1)/{2}}$. Similarly for the restriction of the logarithmic Gelfand-Zeitlin map $\mu: \Herm^+(n)\rightarrow \mathcal{C}(n)$ to $\Herm^+_0(n)$ and $\Sym^+_0(n)$. 

Therefore there is a unique diffeomorphism $\Gamma$ compatible with GZ systems (thus a principal bundle map), called the Alekseev-Meinrenken diffeomorphism, such that for any connected component S of $\Sym_0(n)\subset \Herm(n), \Gamma(S)\subset S$. However, the $\Gamma$ we will construct in the following depend on a parameter space. We expect that they are related to the Alekseev-Meinrenken diffeomorphism via certain isomonodromy flow on the parameter space.

\section{Gelfand-Zeitlin via Stokes phenomenon}\label{defineC}
In this section, let $G$ be a complex reductive Lie group with Lie algebra $\g={\rm Lie}(G)$, and $\mathfrak t\subset\g$ a Cartan subalgebra. Let $\sfPhi\subset\mathfrak t^*$ be the corresponding root system of $\g$.

Let $P$ be the holomorphically trivial principal $G$-bundle on $\mathbb{P}^1$. We consider the following meromorphic connection on $P$ of the form
\begin{eqnarray}\label{nabla}
\nabla:=d-(\frac{\lambda}{z^2}+\frac{1}{2\pi i}\frac{A}{z})dz,
\end{eqnarray}
where $\lambda,A\in\g$. We assume $\lambda\in\mathfrak t$, and once fixed the only variable is $A\in \mathfrak g$. Note that the connection $\nabla$ has an order 2 pole at origin and (if $A\neq 0$) a first order pole at $\infty$. 

\begin{defi}\label{Stokesrays}
The {\it Stokes rays} of the connection $\nabla$ are the rays $\IR_{>0}
\cdot\alpha(\lambda)\subset\mathbb{C}^*$, $\alpha\in\sfPhi$. The {\it Stokes sectors}
are the open regions of $\mathbb{C}^*$ bounded by them.
\end{defi}

Let us choose an arbitrary sector ${\rm Sect}_0$ at $0$ bounded by two adjacent Stokes rays $d_0$, $d_1$, and a branch of ${\rm log}(z)$ on ${\rm Sect}_0$. One fact we will use later is that this sector determines a partition $\sfPhi=\sfPhi_+\cup \sfPhi_-$ of the root system of $\g$. Here $\sfPhi_\pm=\{\alpha\in\sfPhi~|~\alpha (\lambda)\in l, l\in E_\pm\}$, and $E_+$ (resp. $E_-$) is the collection of Stokes rays which one crosses when going from ${\rm Sect}_0$ to $-{\rm Sect}_0$ in the counterclockwise (resp. clockwise) direction.

Now on ${\rm Sect}_0$, there is a canonical
solution $F_0$ of $\nabla$ with prescribed asymptotics on the supersector $\widehat{{\rm Sect}_0}=(d_0-\frac{\pi}{2},d_1+\frac{\pi}{2})$.
In particular, the following result is proved in e.g \cite{BBRS,BJL,MR} for $G={\rm GL}
_n(\mathbb{C})$, in \cite{BoalchG} for $G$ reductive, and in \cite{BTL1} for an
arbitrary affine algebraic group.
Let us denote by $\delta(A)$ the projection of $A$ onto $\mathfrak t$ corresponding to
the root space decomposition $\g=\mathfrak t\bigoplus_{\alpha\in\sfPhi}\g_
\alpha$.

\begin{thm}\label{jurk}
On the sector ${\rm Sect}_0$, there
is a unique holomorphic function $H_0:{\rm Sect}_0\to G$ such that the function
\[F_0=H_0\cdot e^{-\frac{\lambda}{z}}\cdot z^{\frac{\delta(A)}{2\pi i}}\]
satisfies $\nabla F_0=0$, and $H_0$ can be analytically continued to $\widehat{{\rm Sect}_i}$ and then $H_0$ is asymptotic to $1$ within $\widehat{{\rm Sect}_0}$.
\end{thm}


\subsection{Connection matrices}\label{connectiondata}
The meromorphic connection $\nabla=d-(\frac{\lambda}{z^2}+\frac{1}{2\pi i}\frac{A}{z})dz$ is said to be {\it non--resonant}
at $z=\infty$ if the eigenvalues of $\frac{1}{2\pi i}{\rm ad}(A)$ are not positive integers. The following fact is well-known (see e.g \cite{Wasow} for $G={\rm GL}_n(\mathbb{C})$).

\begin{lem}\label{le:nr dkz}
If $\nabla$ is non--resonant, there is a unique holomorphic function
$H_\infty:\mathbb{P}^1\setminus\{0\}\to G$ such that $H_\infty(\infty)=1$, and the function $F_\infty=H_\infty
\cdot z^{\frac{A}{2\pi i}}$ is a solution of $\nabla F=0$.
\end{lem}

Now let us consider the following solutions of $\nabla F=0$:
\begin{eqnarray*}
&&F_0 \ on \ \rm{
Sect}_0,\\
&&F_\infty =H_\infty\cdot z^{\frac{A}{2\pi i}} \ on \ a \ neighbourhood \ of \ \infty \ slit \ along \ d_1,
\end{eqnarray*}
We define the {\bf connection 
matrix} $C(A)\in G$ (with respect to the chosen ${\rm Sect}_0$) by 
\[F_\infty=F_0\cdot C(A).\]
Here $F_\infty$ is extended along a path in ${\rm  Sect}_0$, the identity is understood to hold in the domain of definition of $F_0$. 

Thus we obtain the {\bf connection map} associated to the irregular type $\frac{\lambda}{z}$
\begin{eqnarray}\label{Cmap}
C:\g_{\rm nr}\to G,\end{eqnarray}
which maps any $A\in\mathfrak g_{\rm nr}$ to the connection matrix $C(A)\in G$ of $\nabla$. 
Here $\g_{\rm nr}\subset\g$ is the set of elements $A$
such that the eigenvalues of $\frac{1}{2\pi i}\ad(A)$ do not contain positive integers. 
Assume the centralizer of $\lambda\in\mathfrak t$ is $\mathfrak h=\frak g_1+\frak t\subset \g$ with a semisimple subalgebra $\g_1$, we will denote the connection map $C$ by $C_{\g_1}$. 

\begin{rmk}
Note that the map $C_{\g_1}$ depends not only on $\lambda$, but also on ${\rm Sect}_0$ and a branch of $\log(z)$. Thus when we say a connection map, we will always assume a choice of this data, and the corresponding partition of the root system of $\g$.
\end{rmk}

\subsection{Gelfand-Zeitlin via connection matrices}\label{GZviaconn}
For each $0<k\le n$, we consider the case $\g={\rm gl}_k(\mathbb{C})$. Let $\lambda_k={\rm diag}(a_k,...,a_k,b_k)\in\g$ be a diagonal matrix whose centralizer is the Levi subalgebra $\mathfrak h={\rm gl}_{k-1}(\mathbb{C})+\mathfrak t\subset {\rm gl}_k(\mathbb{C})$. Let $C_k:{\rm gl}_k(\mathbb{C})\rightarrow {\rm GL}_k(\mathbb{C})$ be a (densely defined) connection map associated to the irregular type $\frac{\lambda_k}{z}$ and the Stokes sector $${\rm Sect}_0:=\left\{(a_k-b_k)e^{\iota\phi}~|~\phi\in(-\pi/2,\pi/2)\right\}.$$
\begin{pro}\label{propertyC}
The map $C_k:{\rm gl}_k(\mathbb{C})\rightarrow {\rm GL}_k(\mathbb{C})$ satisfies the conditions
\begin{enumerate}
\item[(a)]  $C_k$ is a ${\rm GL}_{k-1}(\mathbb{C})$-equivariant map, i.e., $C_k(gAg^{-1})={\rm Ad}_gC_k(A)$, for any $g\in {\rm GL}_{k-1}(\mathbb{C})$;
\item[(b)] for any $A\in {\rm gl}_k(\mathbb{C})$, there is a block $\rm LU$ decomposition of $C_k(A) e^A C_k(A)^{-1}$ taking the form $$C_k(A) e^A C_k(A)^{-1}=\left(\begin{array}{cc}
{\rm Id}&0\\
\bar{b}_-& 1
\end{array} \right)\left(\begin{array}{cc}
e^{A^{(k-1)}} & 0\\
0& \star
\end{array} \right)\left(\begin{array}{cc}
{\rm Id} & \bar{b}_+\\
0& 1
\end{array} \right).$$
\end{enumerate}
\end{pro}
\pf Part $(a)$ is straightforward. By definition, the connection matrix $C_k(A)=F_\infty F_0^{-1}$, where $F_0$ and $F_\infty$ are canonical solutions of $\frac{dF}{dz}=(\frac{\lambda_k}{z^2}+\frac{1}{2\pi i}\frac{A}{z})F$ at $0$ and $\infty$ respectively. Set $g\in {\rm GL}_{k-1}(\mathbb{C})\subset {\rm GL}_k(\mathbb{C})$. Due to the fact $g\lambda_k g^{-1}=\lambda_k$, the functions $gF_0g^{-1}$ and $gF_\infty g^{-1}$ are canonical solutions of $\frac{dF}{dz}=(\frac{\lambda_k}{z^2}+\frac{1}{2\pi i}\frac{gAg^{-1}}{z})F$. Thus we have $C_k(gAg^{-1})=gF_\infty F_0^{-1}g^{-1}=gC_k(A)g^{-1}$.

For part $(b)$, let $b_-$, $b_+$ and $e^{A^{(k-1)}}\in {\rm GL}_{k-1}(\mathbb{C})$ be the Stokes matrices and formal monodromy (see e.g. \cite{Boalch5}). A simple fact is that for the Stokes sector ${\rm Sect}_0,$ the matrices $b_+$ ($b_-$) are blocked upper (lower) triangular matrices with diagonal part being identity matrix. Then the identity in $(b)$ is known as the monodromy relation of $\nabla$ $$C(A)e^A C(A)^{-1}=b_-e^{A^{(k-1)}}b_+,$$ translated from the fact that a simple positive loop around $0$ is also a simple negative
loop around $\infty$, where the blocked upper and lower triangular matrices are chosen as $b_\pm$. 
\qed
\begin{lem}\label{restricttoHerm}
For each $0<k\le n$, $C_k:{\rm gl}_k(\mathbb{C})\rightarrow {\rm GL}_k(\mathbb{C})$ restricts to a map (also denoted by) $C_k:\Herm(k)\rightarrow {\rm SU}(k)$. 
\end{lem}
\pf It follows by using the same argument as in (\cite{Boalch1} Lemma 29).\qed\\

Now let $C:=C_1\cdot\cdot\cdot C_n$ be the pointwise multiplication of the connection matrix maps $C_k$ for all $0<k\le n$. Here each $C_k$ is viewed as a map ${\rm gl}_n(\mathbb{C})\rightarrow {\rm GL}_k(\mathbb{C})\subset {\rm GL}_n(\mathbb{C})$ using the projection of ${\rm gL}_n(\mathbb{C})$ onto ${\rm gL}_k(\mathbb{C})$, that is, $C_k(A):=\iota (C_k(A(k)))$. Note that $C_k$ is defined on an open dense subset ${\rm gl}_n(\mathbb{C})_{{\rm nr}^{(k)}}$, consisting of the element $A$ whose $k$th principal submatrix $A^{(k)}$ has no pair of distinct eigenvalues that differ by $2\pi i \mathbb{Z}$. Thus, following the above lemma, $C$ restricts to a map from $\Herm(n)$ to $\Herm^+(n)$. As an immediate consequence of Theorem \eqref{th:GWiso} and Proposition \eqref{propertyC}, we have
\begin{thm}\label{th:GZvStokes}
The map $\Gamma:={\rm Ad_C}\circ{\rm exp}:\Herm(n)\rightarrow\Herm^+(n)$ intertwines the Gelfand-Zeitlin maps and actions.
\end{thm}

\section{Relation to Poisson Lie groups
}\label{RelatoPoisson}
Following \cite{AM}, the linear algebra problem in Section \ref{linearalgebra} can be placed into the context of Poisson
geometry. Consider the Lie algebra ${\rm u}(n)$ of ${\rm U}(n)$, consisting of skew-Hermitian matrices, and identify $\Herm(n)\cong {\rm u}(n)^*$ via the pairing $\langle A,\xi\rangle=2{\rm Im}({\rm tr}A\xi)$. Note that ${\rm u}(n)^*$ carries a canonical linear Poisson structure. On the other hand, the
unitary group ${\rm U}(n)$ carries a standard structure as a Poisson Lie
group (see e.g. \cite{LW1}). The dual Poisson Lie group ${\rm U}(n)^*$, the group of complex
upper triangular matrices with strictly positive diagonal entries,
is identified with $\Herm^+(n)$, by
taking the upper triangular matrix $X\in U(n)^*$ to the positive Hermitian matrix $(X^*X)^{1/2}\in
\Herm^+(n)$. 

These identifications induce densely defined Gelfand-Zeitlin torus actions and maps on ${\rm u}(n)^*$ and ${\rm U}(n)^*$, which are called Gelfand-Zeitlin systems.
It was proved by Guillemin-Sternberg \cite{GS1} that
the action of Gelfand-Zeitlin torus on ${\rm u}(n)^*$ is Hamiltonian, with moment
map the corresponding GZ map $\tau$. As for multiplicative version, Flaschka-Ratiu \cite{FR} proved that the Gelfand-Zeitlin torus action on ${\rm U}(n)^*$ is Hamiltonian, with moment map being the logarithmic GZ
map $\mu$. 

Then our result states that the map $\Gamma$ described in Theorem \ref{th:GZvStokes} actually intertwines the two Hamiltonian systems. That is, given the pointwise multiplication of the connection maps $C:=C_1\cdot\cdot\cdot C_n$, we have
\begin{thm}\label{Poissondiff}
The map $\Gamma={\rm Ad}_C\circ {\rm exp}:{\rm u}(n)^*\rightarrow U(n)^*$ is a Poisson diffeomorphism compatible with the Gelfand-Zeitlin systems.
\end{thm}
\pf In the following sections, we will give more general results. In particular, the proof of this theorem will become clear in Section \ref{compatibleGW}. \qed
\begin{rmk}
Let $C:{\rm gl}_n(\mathbb{C})\rightarrow {\rm GL}_n(\mathbb{C})$ be a connection map associated to an irregular type $\frac{\lambda}{z}$, where $\lambda\in {\rm GL}_n(\mathbb{C})$ is an regular matrix. The remarkable result of Boalch \cite{Boalch1} shows that the irregular Riemann-Hilbert map $\nu:={\rm Ad_C}\circ {\rm exp}$ restricts to a Poisson diffeomorphism ${\rm u}(n)^*\rightarrow {\rm U}(n)^*$. However, as in the above theorem, new relation with Gelfand-Zeitlin systems appears only when we drop the condition of $\lambda$ being regular, and consider a chain of irregular types $\frac{\lambda_k}{z}$ and the corresponding connection maps $C_k$. The discussion above also gives a "moduli theoretic" interpretation of the multiplicative GZ system (i.e. in terms of moduli spaces of meromorphic connections). 
\end{rmk}

\begin{rmk}In \cite{KW}, Kostant and Wallach introduced a
holomorphic Gelfand-Zeitlin system on ${\rm gl}_n(\mathbb{C})$. It is interesting to generalize the above result to that case.
\end{rmk}

In the following sections, we will deepen and find new relations between Stokes phenomenon, symplectic geometry and relative Drinfeld twists, which generalize various results in \cite{AM,Boalch1,Boalch2,EEM,Xu}.

\section{Irregular Riemann-Hilbert correspondence}\label{IrrRHcorrs}
In this section, we will consider a moduli space $\mathcal{A}_{\g_1}$ of (framed) meromorphic connections \eqref{nabla} with an irregular type $\frac{\lambda}{z}$. We assume the centralizer of $\lambda$ is a Levi subalgebra $\mathfrak h=\mathfrak g_1+\mathfrak t$ with semisimple subalgebra $\g_1$ as before. By the irregular Riemann-Hilbert correspondence, this space will be identified with an open dense subset of a space of extended monodromy data $\widetilde{\mathcal{A}}_{\g_1}$, containing the connection matrix and the formal monodromy. $\widetilde{\mathcal{A}}_{\g_1}$ inherits a symplectic structure by the generalized Atiyah-Bott construction \cite{Boalch2, Boalch3}, and $\mathcal{A}_{\g_1}$ is equipped with a natural symplectic structure (see \ref{additivesym}). Then we show that the irregular Riemann-Hilbert map $$\nu:\mathcal{A}_{\g_1}\rightarrow\widetilde{\mathcal{A}}_{\g_1}$$ is a symplectic map. For $\lambda$ regular (thus $\g_1=0$), it has been studied in \cite{Xu} as a special case of \cite{Boalch2, Boalch3}. The generalization to non regular case is direct, thus we will state the main results following the style in \cite{Xu} and ignore some similar proofs.

\subsection{Moduli spaces of (framed) meromorphic connections}
Let $P$ be a holomorphically trivial principal $G$-bundle $P$ over $\mathcal{P}^1$. Let $D=\sum_{i=1}^m k_i(a_i)>0$ be an effective divisor on $\mathbb{P}^1$, and $\nabla$ a meromorphic connection on $P$ with poles on $D$.
In terms of a local coordinate $z$ on $\mathbb{P}^1$ vanishing at $a_i$ and a local trivialisation of $P$, $\nabla$ takes the form of $\nabla=d-A$, where
\begin{eqnarray*}
A=\frac{A_{k_i}}{z^{k_i}}dz+\cdot\cdot\cdot \frac{A_1}{z}dz+A_0dz+\cdot\cdot\cdot,
\end{eqnarray*}
and $A_j\in \g$, $j\le k_i$. We will consider the connection $\nabla$ such that at each $a_i$ the leading coefficient $A_{k_i}\in\g$ is a semisimple element (for $k_i\ge 2$), or the intersection $\g_{\rm reg}\cup \g_{\rm nr}$ (for $k_i=1$). For $G={\rm GL}_n(\mathbb{C})$, those connections $\nabla$ are such that $A_{k_i}$ is diagonalizable with distinct eigenvalues (for $k_i\ge 2$), or diagonalizable with distinct eigenvalues mod $\mathbb{Z}$ (for $k_i=1$).

A {\bf compatible framing} at $a_i$ of $P$ with a connection $\nabla$ is an isomorphism $g_i:P_{a_i}\rightarrow G$ between the fibre $P_{a_i}$ and $G$ such that the leading coefficient of $\nabla$ is inside $\mathfrak t$ in any local trivialisation of $P$ extending $g_0$. We denote by ${\bf g}=\{g_1,...,g_m\}$ the set of compatible framings at all $a_i$.

Let us assume $D=2(a_0)+1(a_1)$, and choose at $a_0$ an irregular type $\frac{\lambda}{z}$,
where $\lambda\in \mathfrak t$. Let $\nabla=d-A$ in some local trivialisation (thus a compatible framing is an element in $G$) and $z$ a local coordinate vanishing at $a_0$. As in \cite{Boalch2}, we say $(\nabla,P)$ with compatible framing $g_0$ at $a_0$ has irregular type $\frac{\lambda}{z}$ if there is some formal bundle automorphism $g\in G\llbracket z\rrbracket$ with $g(a_0)=g_0$, such that $$g[A]:=gAg^{-1}+dg\cdot g^{-1}=\frac{\lambda}{z^2}+\frac{\Lambda}{z}dz$$ for some $\Lambda\in\mathfrak t$.
\begin{defi}
The extended moduli space $\mathcal{A}_{\g_1}$ is the set of isomorphism classes of triples $(P,\nabla,{\bf g})$, consisting of a connection $\nabla$ on $P$ with poles at $D$ and compatible framing ${\bf g}=(g_0,g_1)$, such that $\nabla$ has irregular type $\frac{\lambda}{z}$ at $a_0$.
\end{defi}
Let $\mathfrak t'\subset \mathfrak t$ be the complement of the affine root hyperplanes: $\mathfrak t':= \{t\in \mathfrak t~|~ \alpha(t)\notin 2\pi i \mathbb{Z}\}$. 
\begin{pro}(see e.g. {\rm \cite{Xu})}
The moduli space $\mathcal{A}_{\g_1}$ of the triple $(P,\nabla,{\bf g})$ is isomorphic to $G\times \mathfrak t'$.
\end{pro}
We will introduce a natural symplectic structure on $\mathcal{A}_{\g_1}\cong G\times \mathfrak t'$ in section \ref{Symspace}. In the case $\lambda$ is regular, it has the following origin: if we view $G\times\mathfrak t'$ as a cross-section of $T^*G\cong G\times \g$ (identification via left multiplication and inner product on $\g$), then $G\times \mathfrak t'$ inherits a symplectic structure from the canonical symplectic structure on $T^*G$ (see \cite{GS} Theorem 26.7). Explicitly, it is given as follows.
Let ${\rm ad}^{-1}_x:\mathfrak g\rightarrow \mathfrak g$ be the trivial extension of the map ${\rm ad}^{-1}_x:{\mathfrak g}^\perp_x \rightarrow {\mathfrak g}^\perp_x\subset \mathfrak g$, corresponding to the decomposition $\mathfrak g=\mathfrak g_x\oplus{\mathfrak g}^\perp_x$ for any point $x\in\mathfrak g$. Here $\mathfrak g_x$ is the isotropic subalgebra of $\mathfrak g$ at $x$ and $\mathfrak g_x^\perp$ its complement with respect to the inner product. Then the corresponding Poisson bivector $\pi$ on $G\times \mathfrak t'$ takes the form
\begin{eqnarray}\label{Symslice}
\pi(g,t)=l_g(t_j)\wedge \frac{\partial}{\partial t^j}+l_g({\rm id}\otimes({\rm ad^{-1}_{t}})(\Omega))
\end{eqnarray}
where $\{t_j\}$ is a basis of $\mathfrak t$, $\{t^j\}$ the corresponding coordinates on $\mathfrak t^*$ and $\Omega\in S^2(\g)^\g$ is the Casimir element. Here and in the following, we denote by $l_g$ and $r_g$ the left and right translations by $g$ on $G$.

\subsection{Symplectic spaces of extended monodromy/Stokes data}\label{Stokesdata}
The extended monodromy manifold $\widetilde{\mathcal{A}}_{\g_1}$ is the set of isomorphism classes of Stokes representations of the fundamental groupoid of the irregular curve $\mathbb{P}^1$ with irregular type $D=2(a_0)+1(a_1)$ \cite{Boalch2, Boalch5}.
A quasi-Hamiltonian structure on $\widetilde{\mathcal{A}}_{\g_1}$ can be obtained from an irregular analogue of the Atiyah-Bott construction \cite{Boalch2, Boalch3} via the theory of Lie group valued moment maps \cite{AMM}. It was worked out explicitly in the case $\lambda$ (in the irregular type $\frac{\lambda}{z}$) is regular \cite{Xu}. To remove the assumption, we can follow the strategy in \cite{Boalch4, Boalch5}. However, instead of working in the quasi-Hamiltonian setting, we will work in a Poisson/symplectic geometry setting. We will then point out the equivalence between these two approaches.

Recall that the Lie subalgebra $\mathfrak h\subset \g$ is the centralizer of $\lambda\in\mathfrak t\subset \g$. Then $\lambda$ induces a vector space direct sum $\g={\rm Im(ad_\lambda)}\oplus \mathfrak h$. The subspace ${\rm Im(ad_\lambda)}$ is stabilised by $\mathfrak t$ and can be written as a direct sum of the root spaces of $\g$. On the other hand, the base points of the fundamental groupoid of the irregular curve $\mathbb{P}^1$ determines a choice of positive roots $\sfPhi_+\subset \sfPhi\in \mathfrak t^*$ (similar to the discussion below Definition \ref{Stokesrays}). Set ${\rm Im(ad_\lambda)}=\mathfrak u_+\oplus \mathfrak u_-$ for the subspaces corresponding to positive and negative roots. Then $\mathfrak u_+, \mathfrak u_-\subset \g$ are uilpotent Lie subalgebras, and let $U_+, U_-\subset G$ be the corresponding unipotent subgroups. 

Let $D(G)=G\times G$ be the Heisenberg double equipped with the Poisson tensor $\pi_{D}$ (see \cite{STS}). Thus the manifold $G\times U_-\times U_+\times H$ inherits a Poisson structure via the embedding 
\[\iota: G\times U_-\times U_+\times H\hookrightarrow G\times G; \ (C,u_-,u_+,h)\mapsto (C,u_-^{-1}hu_+).\]
Following \cite{Xu}, let $(G\times \mathfrak t',\pi_\g)$ be the symplectic "slice" of the Lu-Weinstein double symplectic groupoid \cite{LW2} ($\mathfrak t'$ is the complement of the affine root hyperplanes as before).
To describe the space $\widetilde{\mathcal{A}}_{\g_1}$ as a multiplicative symplectic quotient, we consider the product of Poisson spaces
\[\mathcal{M}:=G\times U_-\times U_+\times H\times G\times \mathfrak t',\]
which carries a map \[\mu:M\rightarrow G; \ \mu(g_1,u_-,u_+,h,g_2,t)=g^{-1}_1u_-^{-1}hu_+g_1 g^{-1}_2e^{t} g_2\in G,\] and a $G$ action 
\[g\cdot (g_1,u_-,u_+,h,g_2,t)=(g_1g^{-1},u_-,u_+,h,g_2g^{-1},t).\]

The map $\mu$ and the $G$ action naturally appear in the Stokes representation of the irregular curve $\mathbb{P}^1$ (see \cite{Boalch2, Boalch5} for more details). Then $\widetilde{\mathcal{A}}_{\g_1}$, the set of isomorphism classes of Stokes representations, is isomorphic to $\mu^{-1}(1)/G$.

One checks that the subspace $\mu^{-1}(1)\subset \mathcal{M}$ is coisotropic, and the $G$-invariant functions on it are closed under the Poisson bracket. Thus it induces a symplectic structure on $\widetilde{\mathcal{A}}_{\g_1}\cong \mu^{-1}(1)/G$. The space $\widetilde{\mathcal{A}}_{\g_1}$ equipped with the symplectic structure is called the symplectic space of extended Stokes/monodromy data.
\begin{pro}
The symplectic space $\widetilde{\mathcal{A}}_{\g_1}$ is locally isomorphic to $(G\times \mathfrak t',\pi_{\g})$.
\end{pro}
\pf Explicit formula of $\pi_D$ and $\pi_\g$ can be found in e.g. \cite{Xu}. The proposition then follows from a straightforward computation (see \cite{Xu} Proposition 5.14 for a detailed computation for an additive analogue).
\qed
\\

The Poisson tensor $\pi_\g$ on $G\times \mathfrak t'$ can be expressed by classical $r$-matrices as follows. Let ${r_{\scriptscriptstyle \rm AM}}_\g: \g^*\rightarrow \g\otimes\g$ be the Alekseev-Meinrenken dynamical $r$-matrix \cite{AM0} defined by
\begin{eqnarray*}
{r_{\rm{\scriptscriptstyle AM}}}_\g(x)=({\rm id}\otimes \phi({\rm ad}_{x^{\vee}}))(\Omega), \ \forall x\in\mathfrak g^*,
\end{eqnarray*}
where $x^{\vee}=(x\otimes {\rm id})(\Omega)$ and $\phi(z):=-\frac{1}{z}+\frac{1}{2}{\rm cotanh}\frac{z}{2},$ $z\in \mathbb{C}\setminus 2\pi i\mathbb{Z}^*$. Taking the Taylor expansion of $\phi$ at $0$, we see that $\phi(z)=\frac{z}{12}+\circ(z^2)$, thus $\phi(\rm{ad}_x)$ is well-defined. The maximal domain of definition of $\phi({\rm ad}_x)$ contains all $x\in \mathfrak g^*$ for which the eigenvalues of ${\rm ad}_x$ lie in $\mathbb{C}\setminus 2\pi i \mathbb{Z}^*$.

Let $r_\g$ be the (skewsymmetric part of) standard classical $r$-matrix associated to the partition $\sfPhi=\sfPhi_+\cup\sfPhi_-$. That is $r_\g= \sum_{\alpha\in \sfPhi_+}e_\alpha\wedge e_{-\alpha}$. Then we have
\begin{pro}\rm\cite{Xu}
The Poisson tensor $\pi_\g$ on $G\times \mathfrak t'$ is given by 
$$\pi_{\g}(g,t)=\pi(g,t)+r_g({r_{\scriptscriptstyle \rm AM}}_\g({\rm Ad}_gt))-r_g(r_\g).$$
\end{pro}

{\bf Quai-Hamiltonian $H$-structures on $\widetilde{\mathcal{A}}_{\g_1}$.}
Now we briefly show the quasi-Hamiltonian $H$-structure on the monodromy/Stokes data obtained from the irregular Atiyah-Bott construction. Explicitly, a quasi-Hamiltonian $G\times H$-structure on $G\times U_-\times U_+\times H$ is given in \cite{Boalch4} Theorem 3.1. Thus by taking the fusion product with the quasi-Hamiltonian $G$-space $G\times \t'$ given in \cite{Boalch2} Section 3, one gets a quasi-Hamiltonian $G\times H$ structure on $\mathcal{M}$ with a moment map $\mathcal{M}\rightarrow G\times H$, and the map $\mu:\mathcal{M}\rightarrow G$ (defined above) just takes the first component of this moment map. Eventually, by quasi-Hamiltonian $G$-reduction (see \cite{AMM} Theorem 5.1), we get a quasi-Hamiltonian $H$-structure on $\widetilde{\mathcal{A}}_{\g_1}\cong \mathcal{M}\spr G=\mu^{-1}(1)/G$. 

This quasi-Hamiltonian $H$-structure becomes the Poisson structure $\pi_\g$ on $\widetilde{\mathcal{A}}_{\g_1}$ under the equivalence between various monoidal categories of quasi-Hamiltonian, quasi-Poisson and Poisson $H$-spaces. To be precise, following \cite{AKM} Theorem 10.3, the quasi-Hamiltonian $H$-structure uniquely determines a quasi-Poisson $H$-structure on $\widetilde{\mathcal{A}}_{\g_1}$. This quasi-Poisson $H$-space is further modified by the classical $r$-matrix $r_{\g_1}$ to a Poisson $H$-space, whose Poisson bivector coincides with $\pi_g$. From the perspective of Stokes/monodromy data of meromorphic connections, this procedure amounts to modifying the structure imposed on the formal monodromy in $H$ by Boalch in the quasi-Hamiltonian approach. Thus one can work equivalently in Poisson or quasi-Hamiltonian setting. It may be helpful to compare these two approaches respectively to Fock-Rosly \cite{FoR} and Alekseev-Malkin-Meinrenken \cite{AMM} approaches to the description of the Atiyah-Bott symplectic form on the moduli space of flat connections over Riemann surfaces.

\subsection{Symplectic structure on the moduli space $\mathcal{A}_{\g_1}$}\label{Symspace}
We introduce a (densely defined) symplectic structure on $\mathcal{A}_{\g_1}\cong G\times\mathfrak t'$, which interpolates between the symplectic structures $\pi$ and $\pi_\g$. It is defined as, via the (skewsymmetric part of) standard classical $r$-matrices $r_{\mathfrak g_1}$ and the Alekseev-Meinrenken $r$-matrix $r_{\rm \scriptscriptstyle AM_{\mathfrak g_1}}$ on $\mathfrak g_1$,
\begin{eqnarray}\label{additivesym}
\pi_{\mathfrak g_1}(g,t):=l_g(t_i)\wedge \frac{\partial}{\partial t^i}+l_g(({\rm id}\otimes {\rm ad^{-1}_{t^\vee}})(\Omega))+r_g(r_{\rm \scriptscriptstyle AM_{\mathfrak g_1}}({\rm Ad}_gt))-r_g(r_{\mathfrak g_1}).
\end{eqnarray}
Here $r_{\rm \scriptscriptstyle AM_{\mathfrak g_1}}$ is seen as a function on $\g^*$ via the projection of $\g^*$ onto $\g_1^*$ corresponding to the root space decomposition. The bivector $\pi_{\mathfrak g_1}$ is defined on a dense open subset corresponding to the maximal domain of $r_{\rm \scriptscriptstyle AM_{\mathfrak g_1}}$.
\begin{pro}
The bivector $\pi_{\mathfrak g_1}$ defines a symplectic structure on (a dense subset of) $G\times\mathfrak t'$. 
\end{pro}
\pf Note that at $t\in\g^*$, we have a decompostion $\g=\mathfrak t\oplus \g_t^\perp.$ Assume $\{t_i\}$ is an orthogonal basis of $\mathfrak t$ and $\{f_i\}$ an orthogonal basis of $\g_t^\perp$. 
At each point $(g,t)$, we denote $x:={\rm Ad}_gt\in \g^*$, and $\{{t_i}':={\rm Ad}_g(t_i), {f_i}':={\rm Ad}_g(f_i)\}$ another orthogonal basis of $\g$, $\{{t^i}',{f^i}'\}$ the corresponding coordinates on $\g^*$. 
A straightforward computation of the Schouten-Nijenhuis brackets shows that at $(g,t)$,
$$[l_g(t_i)\wedge \frac{\partial}{\partial t^i},r_g(r_{\rm \scriptscriptstyle AM_{\mathfrak g_1}}(x))]=r_g(t_i'\wedge \frac{\partial r_{\rm \scriptscriptstyle AM_{\g_1}}}{\partial {t^i}'}(x)),$$
and
$$[l_g(({\rm id}\otimes {\rm ad^{-1}_{t}})(\Omega)),r_g(r_{\rm \scriptscriptstyle AM_{\mathfrak g_1}}(x))]=r_g(f_i'\wedge \frac{\partial r_{\rm \scriptscriptstyle AM_{\g_1}}}{\partial {f^i}'}(x)).$$
Therefore, we have $[\pi,r_g(r_{\rm \scriptscriptstyle AM_{\mathfrak g_1}}(x))]=r_g({\rm Alt}( dr_{\rm \scriptscriptstyle AM_{\g_1}}(x)))$. Here recall that $\pi=l_g(t_i)\wedge \frac{\partial}{\partial t^i}+l_g(({\rm id}\otimes {\rm ad^{-1}_{t}})(\Omega))$, and ${\rm Alt}(dr_{\rm \scriptscriptstyle AM_{\mathfrak g_1}}(x))\in \wedge^3\mathfrak g$ is the skew-symmetrization of $dr_{\rm \scriptscriptstyle AM_{\mathfrak g_1}}(x)\in \mathfrak g\otimes\mathfrak g\otimes\mathfrak g$.

On the other hand, due to the fact that $r_{\rm \scriptscriptstyle AM_{\mathfrak g_1}}:\g^*\rightarrow \g\wedge \g$ is $G_1$-equivariant and valued in the subspace $\g_1\wedge \g_1$, we have at point $(g,t)$,
$$[r_g(r_{\rm \scriptscriptstyle AM_{\mathfrak g_1}}(x)),r_g(r_{\rm \scriptscriptstyle AM_{\mathfrak g_1}}(x))]=r_g([r_{\rm \scriptscriptstyle AM_{\mathfrak g_1}}(x),r_{\rm \scriptscriptstyle AM_{\mathfrak g_1}}(x)]).$$ Here the Lie bracket on $\g$ is induced by the left invariant vector fields on $G$, then 
$$[r_g(r_{\mathfrak g_1}),r_g(r_{\mathfrak g_1})]=-r_g(\phi_{\g_1}),$$ where $\phi_{\g_1}\in \wedge^3\g_1\subset \wedge^3\g$ is the Cartan trivector of $\g_1$.  

Eventually, the Schouten-Nijenhuis bracket $$[\pi_{\g_1},\pi_{\g_1}](x)=r_g({\rm Alt}( dr_{\rm \scriptscriptstyle AM_{\g_1}}(x))+[r_{\rm \scriptscriptstyle AM_{\mathfrak g_1}}(x),r_{\rm \scriptscriptstyle AM_{\mathfrak g_1}}(x)])-r_g(\phi_{\g_1}).$$
It is zero because $r_{\rm \scriptscriptstyle AM_{\mathfrak g_1}}$ satisfies the classical dynamical Yang-Baxter equation $${\rm Alt}( dr_{\rm \scriptscriptstyle AM_{\g_1}})+[r_{\rm \scriptscriptstyle AM_{\mathfrak g_1}},r_{\rm \scriptscriptstyle AM_{\mathfrak g_1}}]=\phi_{\g_1}.$$
\qed

The following two examples show that $\pi_{\mathfrak g_1}$ intertwines various known symplectic structures.

\begin{ex}
In the case $\lambda$ is regular (thus $\mathfrak h=\mathfrak t$ and $\g_1=0$), the Poisson space $(G\times \mathfrak t',\pi)$ coincides with the cross-section $G\times \mathfrak t'\subset T^*G\cong G\times \g^*$ with the induced Poisson structure from the canonical symplectic structure on $T^*G$ (see \cite{GS} Theorem 26.7). 
\end{ex}

\begin{ex}
In the case $\mathfrak h=\mathfrak g$, the Poisson space $(G\times \mathfrak t',\pi_\mathfrak g)$ is locally isomorphic to
the symplectic submanifold of Lu-Weinstein double symplectic groupoid studied in \cite{Xu}.
\end{ex}
 
\subsection{Irregular Riemann-Hilbert maps}
Let $(P,\nabla,{\bf g})$ be a triple consisting of a connection $\nabla$ on $P$ with poles at the divisor $D=2(a_0)+1(a_1)$  and compatible framing ${\bf g}=(g_0,g_1)$, such that $\nabla$ has irregular type $\frac{\lambda}{z}$ at $a_0$.
The chosen irregular type canonically determines the Stokes directions at $a_0$, 
and we can consider the Stokes sectors bounded by these directions (and having some small fixed radius). Then the key fact is that, similar to the discussion in section \ref{defineC}, the framings ${\bf g}$ (and a choice of branch of logarithm at each pole) determine, in a canonical way, a choice
of solutions of the equation $\nabla F=0$ on each of the Stokes sectors and on a neighborhood around $a_1$. Then along any path in the punctured sphere $\mathbb{P}^1\setminus \{a_0,a_1\}$ between two Stokes sectors or between one Stokes sector and the neighborhood near $a_1$, we can extend the two corresponding canonical solutions and then obtain a Stokes matrix or a connection matrix valued in $G$ by taking their ratio. The monodromy data of $(P,\nabla,{\bf g})$ is simply the set of all such elements, plus the formal monodromy, thus corresponds to a point in the space of monodromy data $\widetilde{\mathcal{A}}_{\g_1}$. See e.g \cite{Boalch2} Section 3 for more details. 
On the other hand, the moduli space of the triple $(P,\nabla,{\bf g})$ is isomorphic to $\mathcal{A}_{\g_1}$. Therefore, it produces a map from $\mathcal{A}_{\g_1}$ to $\widetilde{\mathcal{A}}_{\g_1}$ by taking the Stokes/monodromy data of meromorphic connections $(P,\nabla, {\bf g})$. 

\begin{thm}\label{RHcorres}
The irregular Riemann-Hilbert map
\begin{eqnarray}\label{monodromymap}
\nu:\mathcal{A}_{\g_1}\rightarrow
\widetilde{\mathcal{A}}_{\g_1}
\end{eqnarray}
associating monodromy/Stokes data to a meromorphic connection $\nabla$ in \eqref{nabla} is symplectic.
\end{thm}
\pf When $\lambda$ in the irregular type $\frac{\lambda}{z}$ of $\nabla$ is regular (thus $\g_1=\frak t$), it becomes a special case of the results in \cite{Boalch3,Boalch4}. For a general $\lambda$, the proof follows a similar way. The only different is that one should also verify the Poisson structure imposed on the formal monodromy $H$ in $\widetilde{\mathcal{A}}_{\g_1}$ coincides under the exponential map with the Poisson structure imposed on the additive analogue $\mathcal{A}_{\g_1}$. This can be seen by a straightforward computation. \qed
\begin{rmk}
We can state a parallel result in the quasi-Hamiltonian setting. For this, we introduce a quasi-Hamiltonian $H$-structure on $\mathcal{A}_{\g_1}\cong G\times t'$ with the corresponding quasi-Poisson (see \cite{AKM}) bivector $\pi_q(g,t):=\pi_{\g_1}+r_g(r_{\mathfrak g_1})$. On the other hand, $\widetilde{\mathcal{A}}_{\g_1}$ is equipped with a quasi-Hamiltonian $H$-structure as in Section \ref{Stokesdata}. Then the Riemann-Hilbert map $\nu:\mathcal{A}_{\g_1}\rightarrow
\widetilde{\mathcal{A}}_{\g_1}$ is a quasi-Hamiltonian map. This is more close to Boalch's origin strategy in \cite{Boalch4,Boalch5}.
\end{rmk}
Now to specify an irregular Riemann-Hilbert map, we have to make a choice of tentacles (see \cite{Boalch2} Definition 3.9), or equivalently a choice of paths generating the fundamental groupoid of the corresponding irregular curve (see \cite{Boalch5}). Using the same choice of tentacles and the same argument as in \cite{Xu}, we have
\begin{pro}\label{RHmap}
The corresponding irregular Riemann-Hilbert map $\nu:\mathcal{A}_{\g_1}\cong (G\times \mathfrak t',\pi_{\g_1})\rightarrow \widetilde{\mathcal{A}}_{\g_1}\cong (G\times \mathfrak t',\pi_\g)$ is given by
\begin{eqnarray*}
\nu(g,t)=(C_{\g_1}({\rm Ad}^*_gt)h, t), \ \forall (g,t)\in G\times \mathfrak t',
\end{eqnarray*}
where $C_{\g_1}$ is the connection map of $\nabla=d-(\frac{\lambda}{z^2}+\frac{1}{2\pi i}\frac{A}{z})dz$.
\end{pro}

As a corollary of Theorem \ref{RHcorres} and Proposition \ref{RHmap}, we have
\begin{thm}\label{Connectioniso}
The map $$\nu_{C_{\g_1}}:(G\times \mathfrak t',\pi_{\g_1})\rightarrow (G\times \mathfrak t',\pi_\g); \ (g,t)=(C_{\g_1}({\rm Ad}^*_gt)g, t)$$
is a local symplectic isomorphism.
\end{thm}
We will unveil the geometric meaning of this theorem in the following section.

\section{Symplectic geometry and dynamical $r$-matrices}\label{Symdynamical}
\subsection{Relative Ginzburg-Weinstein linearization}\label{GWtwist}
Assume we are given the Lie subalgebras $\mathfrak g_1\subset\mathfrak h\subset \g$ with the corresponding Lie groups $G_1\subset H\subset G$ as before. Recall that we have introduced Poisson (symplectic) structures $\pi_\g$ and $\pi_{\g_1}$ on $G\times \mathfrak t'$ associated to this data. We consider the action of $H$ on $G\times \mathfrak t'$, given for any $h\in H$ by
\begin{eqnarray*}
h\cdot (g,t)=(hg,t).
\end{eqnarray*}

\begin{defi}
A relative Ginzburg-Weinstein linearization with respect to $\mathfrak g_1\subset \g$ is an $H$-equivariant locally symplectic diffeomorphism $\Phi_{\mathfrak g_1}:(G\times \mathfrak t',\pi_{\mathfrak g_1})\rightarrow (G\times \mathfrak t',\pi_\g)$, which restricts to the identity map on $G_1\times \mathfrak t'\subset G\times \mathfrak t'$.
\end{defi}
\begin{ex}
The map $\nu_{C_{\g_1}}$ in Theorem \ref{Connectioniso} is $H$-equivariant and restricts to identity map on $G_1\times \mathfrak t'$, due to the $H$-equivariance of the connection map $C_{\g_1}$ and the fact $C_{\g_1}$ restricts to identity map on $\mathfrak h\subset \g$. Thus Theorem \ref{Connectioniso} shows that the map $\nu_{C_{\g_1}}$ is a relative Ginzburg-Weinstein linearization. Another way to construct relative Ginzburg-Weinstein linearization, using the theory of quantum groups, is given in Section \ref{relativetwist}.
\end{ex}
Let $(G\times \mathfrak t',\pi)$ be the symplecitc slice of $T^*G$, where the bivector $\pi$ is given in \eqref{Symslice}.
Note that the $G_1$ action on $G\times \mathfrak t'$ preserves the Poisson structures $\pi_{\g_1}$ and $\pi_\g$. Thus it induces two Poisson algebras on the $G_1$ invariant functions $C^{\infty}(G\times \mathfrak t')^{G_1}$.

\begin{pro}
A relative Ginzburg-Weinstein linearization $\Phi_{\mathfrak g_1}$ with respect to $\mathfrak g_1\subset\mathfrak h\subset \g$ induces a Poisson map from the Poisson algebra $(C^{\infty}(G\times \mathfrak t')^{G_1},\pi_\g)$ to $(C^{\infty}(G\times \mathfrak t')^{G_1},\pi)$.
\end{pro}
\pf 
The bivector field $(\pi_{\mathfrak g_1}-\pi)(g,t)=r_g(r_{\rm \scriptscriptstyle AM_{\mathfrak g_1}}({\rm Ad}_gt))-r_g(r_{\mathfrak g_1})$ vanishes on $G_1$-invariant functions $C^\infty(G\times \mathfrak t)^{G_1}$, that is because $r_{\rm \scriptscriptstyle AM_{\mathfrak g_1}}$ and $r_{\g_1}$ are valued in $\g_1\wedge \g_1$. Thus $\pi_{\mathfrak g_1}$ coincides with $\pi$ while restricting to the $G_1$ invariant functions. The proposition follows immediately.\qed\\

In the case $\mathfrak g_1=0$ (thus $\mathfrak h=\mathfrak t$), we will show that the above proposition recovers the Ginzburg-Weinstein linearization for the dual Poisson Lie group $(G^*,\pi_{G^*})$ associated to the standard classical r-matrix $r_\g$. For this, let us consider the Semenov-Tian-Shansky (STS) Poisson bivector on $\mathfrak g$,
\begin{eqnarray*}
\pi_{\rm STS}(x)(df,dg)=\langle df(x)\otimes dg(x), {\rm ad}_x\otimes\frac{1}{2}{\rm ad}_x{\rm coth}(\frac{1}{2}{\rm ad}_x)(\Omega)-\otimes^2{\rm ad}_x(r_\g)\rangle,
\end{eqnarray*}
for any $f, g\in C^{\infty}(\g)$. By the definition of $\Phi_{\g_1=0}$, we have the following commutative diagram of Poisson maps
$$
\begin{CD}
(G\times \mathfrak t',\pi) @> \Phi_{0}>> (G\times \mathfrak t',\pi_{\g}) \\
@V P VV @V P VV \\
(\mathfrak g^*,\pi_{\rm \scriptscriptstyle KKS} )@> \Phi_0' >> (\mathfrak g,\pi_{\rm\scriptscriptstyle STS})
\end{CD},
$$
where $P:G\times \mathfrak t'\rightarrow \g^*; \ (g,t)\mapsto {\rm Ad}_gt$ is the projection with respect to the $H=T$ action on $G\times \mathfrak t'$, and $\Phi_0'$ is the induced Poisson map. Following \cite{FM}, the map $(\mathfrak g,\pi_{STS})\rightarrow (G^*,\pi_{G^*}); \ x\mapsto (b_-(x),b_+(x))$, determined by the decompostion $e^{x}=b_-(x)^{-1}b_+(x)$, is a local Poisson isomorphism. Therefore in this case, the linearization $\Phi_{0}$ reduces to a Poisson algebra map $C^{\infty}(G^*)\rightarrow C^{\infty}(\g^*)$.

\subsection{Relative gauge transformation equations between $r$-matrices}\label{sectiongauge}
Let ${r_{\scriptscriptstyle \rm AM}}_\g: \g^*\rightarrow \g\otimes\g$ (resp ${r_{\scriptscriptstyle \rm AM}}_{\g_1}:\g_1\rightarrow \g_1\wedge\g_1$) be the Alekseev-Meinrenken dynamical $r$-matrix for $\g$ (resp. $\g_1$). We view ${r_{\scriptscriptstyle \rm AM}}_{\g_1}$ as a function on $\g^*$ (valued in $\g_1\wedge\g_1\subset\g\wedge\g$) via the projection of $\g\cong\g^*$ onto $\g_1\cong\g_1^*$.

\begin{defi}\label{relativeeq}
The relative gauge transformation equation for a map $\rho\in {\rm Map}(\g^*,G)$ is (as identity of maps $\g^*\rightarrow \wedge^2(\g)$)
\begin{eqnarray}\label{gaugeequation}
{r_{\scriptscriptstyle \rm AM}}_\g-(\otimes^2 {\rm Ad}_{\rho^{-1}})(r_{{\scriptscriptstyle \rm AM}_{\g_1}})=(r_\g-r_{\g_1})^\rho.
\end{eqnarray}
Here $$(r_\g-r_{\g_1})^\rho:= \rho_1^{-1}d_2(\rho_1)-\rho_2^{-1}d_1(\rho_2)+(\otimes^2{\rm Ad}_\rho)^{-1}(r_\g-r_{\g_1})+\langle {\rm id}\otimes {\rm id}\otimes x,[\rho_1^{-1}d_3(\rho_1),\rho_2^{-1}d_3(\rho_2)]\rangle,$$ and $\rho_1^{-1}d_2(\rho)(x)=\sum_i \rho^{-1}\frac{\partial \rho}{\partial {\xi^{i}}}(x)\otimes e_i$ is viewed as a formal
function $\mathfrak g^*\rightarrow \mathfrak g^{\otimes 2}$, $\{e_i\}$ is a basis of $\mathfrak g$, $\{\xi^i\}$ the corresponding coordinates on $\mathfrak g^*$ and $\rho_i^{-1}d_j(\rho_i) = (\rho_1^{-1}d_2(\rho_1))^{i,j}$.
\end{defi}

\begin{rmk}
In \cite{EEM}, Enriquez, Etingof and Marshall introduced the gauge transformation equation for a map $\rho\in {\rm Map}(\g^*,G)$
\begin{eqnarray}\label{Geq}
r_\g^\rho=r_{\rm{\scriptscriptstyle AM}}. 
\end{eqnarray}
Associated to a formal solution $\rho:\mathfrak g^*\rightarrow G$ of \eqref{Geq}, they constructed formal Poisson isomorphisms between the formal Poisson manifolds $\mathfrak g^*$ and $G^*$. They also derived this equation as a semiclassical limit of the vertex-IRF gauge transformation between dynamical twists \cite{EN}.
\end{rmk}

\begin{defi}
A solution $\rho\in {\rm Map}(\g^*,G)$ of \eqref{gaugeequation} is called a relative Ginzburg-Weinstein twist if $\rho$ is $\mathfrak h$-equivariant, and $\rho(x)=1$ for any $x\in \g^*_1\subset \g^*$.
\end{defi}

\subsection{Linearization by relative Ginzburg-Weinstein twists}

Given a map $\rho:\g^*\rightarrow G$, we define a diffeomorphism $\nu_\rho:G\times \mathfrak t'\rightarrow G\times \mathfrak t'$ by 
\begin{eqnarray}
\nu_{\rho}(g,t)=(\rho({\rm Ad}^*_g(t))g,t).
\end{eqnarray}

The symplectic geometric interpretation of the relative gauge equation \eqref{gaugeequation} is then given in the following proposition.
\begin{pro}\label{relatsymplectic}
Given the Lie subalgebras $\mathfrak g_1\subset \mathfrak h\subset \g$, a map $\rho\in{\rm Map}(\g^*,G)$ is a relative Ginzburg-Weinstein twist if and only if $\nu_\rho:(G\times\mathfrak t',\pi_{\mathfrak g_1})\rightarrow (G\times \mathfrak t',\pi_\g)$ is a relative Ginzburg-Weinstein linearization.
\end{pro}
\pf First, given an $H$-equivariant map $\rho\in {\rm Map}(\g^*,G)$, we show that the following two conditions are equivalent:

$(a)$ The map $\rho\in {\rm Map}(\g^*,G)$ satisfies the relative gauge equation \eqref{gaugeequation}.

$(b)$ The diffeomorphism $\nu_{\rho}:G\times \mathfrak t'\rightarrow G\times \mathfrak t'; \ (g,t)
\mapsto (\rho({\rm Ad}^*_g(t))g,t)$
intertwines the Poisson structure $\pi_{\mathfrak g_1}$ and $\pi_\g$.

Let us first compute ${\nu_{\rho}}_*(\pi_{\g_1})$, where $\pi_{\g_1}(g,t)=\pi+r_g(r_{\rm \scriptscriptstyle AM_{\mathfrak g_1}}({\rm Ad}_gt))-r_g(r_{\mathfrak g_1})$.
At each point $(g,t)\in G\times\mathfrak t'$, set $x:={\rm Ad}^*_ht \in\mathfrak g^*$, then $\nu_{\rho}(g,t)=(\rho(x)g,t)$. 

We take $\{e^i\}$, $\{e_i\}$ as dual bases of $\mathfrak g^*$, $\mathfrak g$ and $\{t^j\}$, $\{t_j\}$ dual bases of $\mathfrak t^*$, $\mathfrak t$. A straightforward calculation gives that at each point $(\rho(x)g,t)\in G\times \mathfrak t'$
\begin{eqnarray*}
{\nu_{\rho}}_*(l_g(e_i))&=&l_{\rho g}(e_i)+l_{\rho g}(g^{-1}\rho ^{-1}\frac{\partial \rho }{\partial X^i}g),\\
{\nu_{\rho}}_*(\frac{\partial}{\partial t_j})&=&\frac{\partial}{\partial t_j}+l_{\rho g}(g^{-1}\rho^{-1}\frac{\partial \rho}{\partial T^j}g)
\end{eqnarray*}
where $X^i:=[{\rm Ad}_ge_i,x]$, $T^j:={\rm Ad}^*_gt^j$ are tangent vectors at $x={\rm Ad}^*_gt$. Note that $T^m\in\mathfrak g_x$ (the isotropic subalgebra at $x$) and $X^i$ span the tangent space $T_x\mathfrak g^*$ and thus the above formulas involve all the possible derivative of $\rho\in {\rm Map}(\mathfrak g^*,G)$.
A direct computation shows that at each point $(\rho(x)g,t)\in G\times\mathfrak t'$ (here $x={\rm Ad}^*_gt\in\mathfrak g^*$)
\begin{eqnarray*}
{\nu_{\rho}}_*(\pi)(\rho(x)g,t)=\pi(\rho(x)g,t)+l_{\rho g}(\otimes^2{\rm Ad}_{g^{-1}}U(x)),
\end{eqnarray*}
where $U(x)\in \mathfrak g\wedge \mathfrak g$ is defined as
\begin{eqnarray*}
U(x)=\rho_1^{-1}d_2(\rho_1)-\rho_2^{-1}d_1(\rho_2)+\langle {\rm id}\otimes {\rm id}\otimes x,[\rho_1^{-1}d_3(\rho_1),\rho_2^{-1}d_3(\rho_2)]\rangle.
\end{eqnarray*}

On the other hand, because $r_{\rm \scriptscriptstyle AM_{\mathfrak g_1}}(x)$ is valued in $\g_1\wedge \g_1\subset \g\wedge \g$ and the map $\rho$ is $H$-equivariant, the pushforward of the bivector $r_g(r_{\rm \scriptscriptstyle AM_{\mathfrak g_1}}(x))$ by $\nu_\rho$ is
$${\nu_{\rho}}_*(r_g(r_{\rm \scriptscriptstyle AM_{\mathfrak g_1}}(x)))=r_{\rho g}(r_{\rm \scriptscriptstyle AM_{\mathfrak g_1}}(x)).$$

Thus by comparing with the expression of $\pi_\g$,
\begin{eqnarray*}
\pi_\g(\rho(x)g,t)=\pi(\rho(x)g,t)+l_{\rho g}(r_{\rm \scriptscriptstyle AM}(t)-\otimes^2{\rm Ad}_{(\rho g)^{-1}}r_0),
\end{eqnarray*}
we obtain that ${\nu_{\rho}}_*(\pi_{\g_1})=\pi_\g$ at point $(\rho(x)g,t)\in G\times \mathfrak t'$ if and only if
\begin{eqnarray*}
r_{\rm \scriptscriptstyle AM}(t)-{\rm Ad}_{(\rho g)^{-1}}r_{\rm \scriptscriptstyle AM_{\mathfrak g_1}}(x)=\otimes^2{\rm Ad}_{(\rho g)^{-1}}r_0+\otimes^2Ad_{g^{-1}}U(x).
\end{eqnarray*}
Note that $x={\rm Ad}^*_gt$, by the $G$-equivariance of $r_{\rm \scriptscriptstyle AM}$, we have $\otimes^2{\rm Ad}_gr_{\rm \scriptscriptstyle AM}(t)=r_{\rm \scriptscriptstyle AM}(x)$. Thus the above identity is exactly the gauge transformation equation \eqref{gaugeequation}.

Furthermore, we have
\begin{itemize}
\item the map $\rho:\g^*\rightarrow G$ is $H$-equivariant if and only if $\nu_\rho$ is $H$-equivariant;

\item $\rho(x)=1$ for any $x\in \g^*_1\subset\g^*$ if and only if $\nu_\rho$ restricts to the identity map on $G_1\times \mathfrak t^*\subset G\times \mathfrak t'$.
\end{itemize}

It follows that $\rho$ is a Ginzburg-Weinstein twist ($H$-equivariant, restricts to $1$ on $\g_1^*$, and satisfies equation \eqref{gaugeequation}), if and only if the map $\nu_\rho:(G\times \mathfrak t',\pi)\rightarrow (G\times \mathfrak t',\pi_\g)$ is a relative Ginzburg-Weinstein linearization. \qed

\subsection{Connection maps as relative Ginzburg-Weinstein twists}
As a corollary of Theorem \ref{Connectioniso} and Proposition \ref{relatsymplectic}, we have
\begin{thm}\label{main}
The connection map $C_{\g_1}:\g^*\to G$ satisfies the gauge equation \eqref{gaugeequation}, i.e., $(r_\g-r_{\g_1})^{C_{\g_1}}=r_{\rm \scriptscriptstyle AM}(x)$.
\end{thm}

\subsection{Composition of gauge transformations}
Assume we are given a relative Ginzburg-Weinstein twist $C_{\g_1}\in {\rm Map}(\g^*,G)$ with respect to $\g_1\subset \g$. Let $G_1\subset G$ be the integration of $\g_1$. 
\begin{pro}\label{twistcomposition0}
A map $\rho_{\g_1}\in {\rm Map}(\g_1^*,G_1)$ satisfies $r_{\g_1}^{\rho_{\g_1}}={r_{\rm\scriptscriptstyle AM}}_{\g_1}$ if and only if the pointwise mutiplication $\rho:=\rho_{\g_1}\cdot C_{\g}\in {\rm Map}(\g^*,G)$ satisfies $r_{\g}^\rho={r_{\scriptscriptstyle \rm AM}}_{\g}$ (provided $\rho_{\g_1}$ is seen as a map from $\g^*$ to $G$ via the projection of $\g^*$ onto $\g_1^*$ and the inclusion of $G_1$ to $G$).
\end{pro}
\pf Note that the composition of diffeomorphisms
$$\nu:(G\times \mathfrak t',\pi) \stackrel{\nu_{\rho_{\g_1}}}\longrightarrow (G\times \mathfrak t',\pi_{\g_{1}})\stackrel{\nu_{C_\g}}\longrightarrow(G\times\mathfrak t',\pi_{\g})$$
is still a diffeomorphism. We show that $\nu:=\nu_{C_\g}\circ \nu_{\phi_{\g_1}}$ coincides with $\nu_{\rho}$. This is because $\rho_{\g_1}$ is valued in $G_1\subset G$ and $C_\g$ is $G_1$-equivariant, then we have $$\nu(g,t)=\nu_{C_\g}\circ \nu_{\phi_{\g_1}}(g,t)=(\phi_{\g_1}(gt g^{-1})C_\g(gt g^{-1})g,t)=\nu_{\phi_{\g_1}\cdot C_\g}(g,t).$$
Therefore, given $\nu_{C_\g}$ is symplectic, the map $\nu_\rho=\nu_{C_\g}\circ \nu_{\rho_{\g_1}}$ is symplectic if and only if $\nu_{\rho_{\g_1}}$ is. By Proposition \ref{relatsymplectic}, this finishes the proof.
\qed\\

Let $G$ (resp. $G_1$) be equipped with the Poisson Lie group structure associated to the quasi-triangular Lie bialgebra $(\g,r_\g)$ (resp. $(\g_1,r_{\g_1}))$. Then the Lie group homomorphism (inclusion) $\mathcal{T}_1:G_1\rightarrow G$ is a Poisson Lie group homomorphism. We denote by $\tau_1:\g_1\rightarrow \g$ the corresponding infinitesimal Lie algebra morphism, $\mathcal{T}_1^*:G^*\rightarrow G^*_1$ the dual Poisson Lie group morphism.

Following \cite{EEM} Proposition 0.3, associated to a solution of $r_{\g}^\rho={r_{\scriptscriptstyle \rm AM}}_{\g}$, there is a unique densely defined Poisson isomorphims $\Gamma_{\rho}:\g^*\rightarrow G^*$.
Thus Propostion \eqref{twistcomposition0} gives the following commutative diagram
\[
\begin{CD}
\g^* @>{\tau_{1}^*}>>\g_{1}^*\\
@VV{\Gamma_{{\scriptscriptstyle {\rho_{\g_1}}{ C_{\g_1}}}}}V     @VV{\Gamma_{\rho_{\g_{1}}}}V \\
G^* @>>{{\mathcal{T}}_{1}^*}>G_{1}^*
\end{CD}
\]

\begin{rmk}
In \cite{AM1} Theorem 4.1, Alekseev and Meinrenken proved the functorial property of Ginzburg-Weinstein linearization for coboundary Poisson Lie groups. Their result states that given any $\rho_{\g_1}$, there exists a Ginzburg-Weinstein twist $\rho$ such that the above diagram (with $\Gamma_{\rho}$ replacing $\Gamma_{{\scriptscriptstyle {\rho_{\g_1}}{ C_{\g_1}}}}$) commutes. Thus in the case $G$ is a standard Poisson Lie group, we have explicitly constructed $\rho$ in a universal way via a connection map $C_{\g_1}$. In this sense, $\Gamma_{\scriptscriptstyle C_{\g_1}}$ can be understood as a "universal" relative linearization of the dual Poisson Lie group $G^*$ (relative to the subgroup $G_1^*$).
\end{rmk}

\subsection{Ginzburg-Weinstein linearizations compatible with nested sets}\label{compatibleGW}
A nested set on the Dynkin diagram $D$ of $\g$ is a collection of pairwise compatible, connected subdiagrams of $D$ containing $D$. 
For example, If $D$ is the Dynkin diagram of type $A_{n-1}$, with vertices labelled $1,...,n-1$, nested sets on $D$ are in bijection with bracketings of the non associative monomial $x_1\cdot\cdot\cdot x_n$. One example of maximal nested set is $(\cdot\cdot\cdot((x_1)x_2)\cdot\cdot\cdot x_{n-1})$.

Fix a maximal nested set on $D$, with the subdiagram $D_0=\emptyset\subset \cdot\cdot\cdot\subset D_{n-1}\subset D_n=D$.
For any subdiagram $D_i\subset D$, let $\g_i\subset \g$ be the subalgebra generated by the root subspaces $\g_{\pm\alpha_j}$, $\alpha_j\in D_i$. Thus we get a chain $\g_0=0\subset\g_1\cdot\cdot\cdot \g_{n-1}\subset \g_n=\g$ of Lie sublagebras of $\g$.

Let $r_{\g_i}$ be the skewsymmetric part of the standard $r$-matrix of $\g_i$ with respect to a positive root system. By Theorem \ref{main}, for each $i$, we can define a relative twist $C_{i}\in {\rm Map}(\g_i^*,G_i)$ with respect to the pair $\g_{i-1}\subset \g_{i}$ via Stokes phenomenon. Then we can prove inductively that 

\begin{pro}\label{comptwist}
$\rho_i:=C_1\cdot\cdot\cdot C_{i}\in {\rm Map}(\g^*_i,G_i)$ satisfies the equation \eqref{Geq} for $\g_i$, i.e., $r_{\g_i}^{\rho_i}={r_{\scriptscriptstyle \rm AM}}_{\g_i}$. Here for any $j<i$, $C_j$ is seen as a map from $\g^*_i$ to $G_i$ (via the projection of $\g^*_i$ onto $\g_j^*$ and the inclusion $G_j\rightarrow G_i$). 
\end{pro}
Recall that we have $C_i(x)=1$, for any $x\in\g^*_{i-1}\cong \g_{i-1}\subset \g^*_i\cong \g_i$. Geometrically, the above proposition reflects the fact that the composition of symplectic maps
$$\nu_{\rho_i}:(G\times\mathfrak t',\pi_{\g_{0}}=\pi) \stackrel{\nu_{C_1}}\longrightarrow (G\times\mathfrak t',\pi_{\g_{1}})\longrightarrow\cdot\cdot\cdot \longrightarrow (G\times\mathfrak t',\pi_{\g_{i-1}})\stackrel{\nu_{C_{i}}}\longrightarrow (G\times\mathfrak t',\pi_{\g_i}),$$
is symplectic.

Denote by $G_i$ the simply connected Lie group equipped with Poisson Lie group structure associated to the quasi-triangular Lie bialgebra $(\g_i,r_{\g_i})$. The Lie group morphism (inclusion) $\mathcal{T}_{i-1}:G_{i-1}\rightarrow G_i$ is a Poisson Lie group morphism. We denote by $\tau_{i-1}:\g_{i-1}\rightarrow \g_i$ the corresponding infinitesimal Lie algebra morphism, by $\mathcal{T}_{i-1}^*:G^*_i\rightarrow G^*_{i-1}$ the dual Poisson Lie group morphism.

Then those $\rho_i$ in Proposition \eqref{comptwist} are compatible in the sense that the resulting diagram commutes

\[\label{eq:diagram0}
\begin{CD}
\g^* @>{\tau_{n-1}^*}>>\g_{n-1}^* @>{\tau_{n-2}^*}>> \cdots @>{\tau_{1}^*}>>  \g_{1}^*\\
@VV{\Gamma}V     @VV{\Gamma_{\rho_{n-1}}}V     @.            @VV{\Gamma_{\rho_{1}}}V \\
G^* @>>{{\mathcal{T}}_{n-1}^*}>G_{n-1}^* @>> {{\mathcal{T}}^*_{n-2}}> \cdots @>>{{\mathcal{T}}_{1}^*}> G_{1}^* 
\end{CD}
\]
where each $\Gamma_{\rho_i}:\g_i^*\rightarrow G^*_i$ is the linearization given by the $\rho_i$ as in (\cite{EEM} Proposition 0.3). Here $\g=\g_n$ and $\Gamma:=\Gamma_{\rho_n}$.
\\
\\
{\bf The $U(n)$ Gelfand-Zeitlin system.}
Consider the special case $\g={\rm gl}_n(\mathbb{C})$. Let $C$ be the multiplication of the connection maps $C_i$ as in Section \ref{GZviaconn}. Then Theorem \ref{th:GZvStokes} and the above discussion give a proof of Theorem \ref{Poissondiff}.

\section{Semiclassical limit of relative Drinfeld twists}\label{relativetwist}
Let $(U(\g),m,\Delta,\varepsilon)$ denote the universal enveloping algebra of $\g$ with the product $m$, the coproduct $\Delta$ and the counit $\varepsilon$. Let $U(\g)\llbracket\hbar\rrbracket$ be the corresponding topologically free $\mathbb{C}\llbracket\hbar\rrbracket$-algebra. 

Set $U:=U(\g)\llbracket\hbar\rrbracket$ and $U':=U(\hbar \g)\llbracket\hbar\rrbracket$, the subalgebra generated by $\hbar x$, $\forall x\in \g$. Note that $U'/\hbar U'=\hat{S}(\g)$. An associator $\Phi\in U(\g)^{\widehat{\otimes}3}\llbracket\hbar\rrbracket$ is called admissible (see \cite{EH}) if
$$\Phi\in 1+\frac{\hbar^2}{24}[\Omega^{1,2},\Omega^{2,3}]+O(\hbar^3),\ \ \ \ \ \hbar {\rm log}(\Phi)\in U'^{\widehat{\otimes}3}.$$

As before, let us take $\g_1\subset\mathfrak h\subset\g$.
Let $\Phi_g$ (resp. $\Phi_{\g_1}$) be an admissible associator of $\g$ (resp. $\g_1$). 
Then an $\mathfrak h$-invariant element $J\in (U(\g)^{\otimes{2}}\llbracket\hbar\rrbracket)^\mathfrak h$ is a relative twist if it satisfies the equation
\begin{equation} \label{J:Phi}
\Phi_\g = (J^{2,3}J^{1,23})^{-1}\Phi_{\g_1} J^{1,2} J^{12,3}. 
\end{equation}

Let $J$ be an admissible relative twist quantization of the relative classical $r$-matrix $r_{\g}-r_{\g_1}$. That is $J$ is a relative twist, and satisfies $J=1-\hbar\frac{r_{\g}-r_{\g_1}}{2}+\circ(\hbar)$, $\hbar {\rm log}(J)\in U'^{\widehat{\otimes}2}, (\varepsilon\otimes {\rm id})(J)=({\rm id} \otimes \varepsilon)(J)=1$,

We identify the second component $U(\g)$ of $J$ with $\mathbb{C}[\g^*]$ via the symmetrization (PBW) isomorphism $\hat{S}(\g)\rightarrow U(\g)$, and thus we get a formal function from $\g^*$ to $U(\g)\llbracket\hbar\rrbracket$, denoted by $J(x)$. 
We have ${\rm Ker}(\eps) \cap U' \subset \hbar U$, therefore 
$J\in U \wh\otimes U'$. Thus $J(\hbar^{-1}x):\g^*\rightarrow U(\g)\llbracket\hbar\rrbracket$ is well-defined. Denote by $g(x)
= J(\hbar^{-1}x)_{|\hbar=0}$ its reduction mod $\hbar$, 
which is a formal series on $\g^*$ with coefficients in $U(\g)$. 
One checks that the reduction mod $\hbar$ of (\ref{J:Phi}) is 
$g^{12}(x) = g^{1}(x)g^{2}(x)$. Since we also have $(\eps\otimes \id)(g)=1$, 
we get that $g(x)$ is a formal series on $\g^*$ with coefficients in the formal group ${\rm exp}(\g)$. 

\begin{pro} \label{pro:relativeeq}
The map $g(x)$ is a formal solution of the equation \eqref{gaugeequation}. 
\end{pro}

\pf According to \cite{EE}, $\Phi_\g\in U^{\wh\otimes 2}
\wh\otimes U'$ has the expansion $1+\hbar \phi_\g + o(\hbar^2)$, 
where $\phi_\g \in U^{\wh\otimes 2} \wh\otimes U'$ is such that 
$(\phi_\g - \phi_\g^{2,1,3})_{|\hbar=0} = -{r_{\scriptscriptstyle \rm AM}}_{\g}\in U^{\wh\otimes 2}
\wh\otimes \wh S(\g)$. Similarly, $\Phi_{\g_1}=1+\hbar \phi_{\g_1} + o(\hbar^2)$, with $(\phi_{\g_1}- \phi_{\g_1}^{2,1,3})_{|\hbar=0} = -{r_{\scriptscriptstyle \rm AM}}_{\g_1}\in U^{\wh\otimes 2}
\wh\otimes \wh S(\g)$. Recall that ${r_{\scriptscriptstyle \rm AM}}_{\g_1}$ is seen as a function on $\g$ via the projection of $\g$ onto $\g_1$.

In the following, given any $f\in U(\g)\wh\otimes \wh S(\g)$, we denote by $\bar{f}$ a lift of $f$ in 
$U\wh\otimes U'$. 
Because $g$ is a formal series on $\g^*$ valued in ${\rm exp}(\g)$, we can expand  $\log(J)=\overline{\log(g)} + \hbar X + o(\hbar)$, with $X\in 
U\wh\otimes U'$. 
Then $J^{1,23} = \exp(\overline{\log(g)^{1,3}}+ \hbar(X^{1,3} + \overline{d_2 \log(g)^{1,3}}) 
+ o(\hbar)) = J^{1,3} (1 + \hbar \overline{g_1^{-1}d_2(g_1)} + o(\hbar))$. Here we take the notation in Definition \ref{relativeeq}.

On the other hand, $[J^{1,3},J^{2,3}] = \hbar \overline{\{g^{1,3},g^{2,3}\}} + o(\hbar)$, 
so $$(J^{12,3})^{-1}[J^{1,3},J^{2,3}] = \hbar \overline { 
(g^{1,3}g^{2,3})^{-1}\{g^{1,3},g^{2,3}\} } + o(\hbar) = 
\hbar \overline{ \langle \id\otimes \id\otimes x, 
[g_1^{-1}d_3(g_1),g_2^{-1}d_3(g_2)] \rangle} + o(\hbar).$$ 
Thus we get 
$$
J^{12,3} = J^{2,3} J^{1,3} (1 + \hbar Y + o(\hbar)), 
$$
where $Y \in U^{\wh\otimes 2}\wh\otimes U'$ is such that 
$(Y - Y^{2,1,3})_{|\hbar=0} = \langle \id\otimes \id\otimes x,
[g_1^{-1}d_3(g_1),g_2^{-1}d_3(g_2)] \rangle$. 

Then (\ref{J:Phi}) gives 
\begin{eqnarray*}
1 + \hbar \phi_\g + o(\hbar)&=& (1 - \hbar \overline{g_1^{-1} d_2(g_1)} 
+ o(\hbar))
(J^{1,3})^{-1} (J^{2,3})^{-1}\\&& \cdot (1+\phi_{\g_1}+o(\hbar))(1 - \hbar\frac{r_\g-r_{\g_1}}{2} + o(\hbar)) J^{2,3}
J^{1,3} (1 + \hbar Y + o(\hbar)). 
\end{eqnarray*}
The reduction modulo $\hbar$ of $(J^{1,3})^{-1} (J^{2,3})^{-1} (r_\g-r_{\g_1}) 
J^{2,3} J^{1,3}$ is $\Ad(g\otimes g)^{-1}(r_\g-r_{\g_1})$. Similarly, the reduction modulo $\hbar$ of $(J^{1,3})^{-1} (J^{2,3})^{-1} (\phi_{\g_1}-\phi_{\g_1}^{2,1,3}) 
J^{2,3} J^{1,3}$ is $\Ad(g\otimes g)^{-1}({r_{\scriptscriptstyle \rm AM}}_{\g_1}).$ Then the proposition follows by substracting $1$, 
dividing by $\hbar$, reducing modulo $\hbar$ and 
antisymmetrizing the two first tensor factors.
\qed\\

Due to the $\frak h$-invariance of the relative twist $J$, the semiclassical limit $g(x)$ of $J$ is an $\frak h$-equivariant map. Therefore, it gives rise to a formal relative Ginzburg-Weinstein twist (provided $g(x)$ restricts to the identity map on $\g_1^*$), and thus (by Proposition \ref{relatsymplectic}) to a relative linearization.

Because the connection map $C_{\g_1}$ is also a solution of \eqref{gaugeequation}, one natural question is that if there exists a relative Drinfeld twist whose semiclassical limit is $C_{\g_1}$. Such a relative twist may be constructed as a (quantum) connection matrix of the dynmaical Knizhnik–Zamolodchikov equation \cite{FMTV} by slightly generalizing the construction in \cite{TL} to allow for non-regular $\lambda$, and then can be shown to be a quantization of the map $C_{\g_1}$ following the idea from \cite{TLXu}.

\Addresses

\end{document}